\documentclass[lettersize,journal]{IEEEtran}
\usepackage{amsmath,amsfonts}
\usepackage{algorithmic}
\usepackage{algorithm}
\usepackage{array}
\usepackage[caption=false,font=normalsize,labelfont=sf,textfont=sf]{subfig}
\usepackage{textcomp}
\usepackage{stfloats}
\usepackage{url}
\usepackage{verbatim}
\usepackage{graphicx}
\usepackage{cite}
\usepackage[hidelinks]{hyperref}
\usepackage{amsthm}
\usepackage{fmtcount}

\newcommand{\R}{\mathbb{R}}

\newtheorem{assumption}{Assumption}[section] 

\newtheorem{problem}{Problem}[section]

\newtheorem{remark}{Remark}

\newtheorem{theorem}{Theorem}[section]

\newcommand\norm[1]{\lVert#1\rVert} 

\begin{document}

\title{Vision-Based Control For Landing an Aerial Vehicle on a Marine Vessel}

\author{Haohua Dong, David Cabecinhas and Rita Cunha

Institute for Systems and Robotics

Instituto Superior Técnico, University of Lisbon, Portugal

\thanks{Extended Abstract}}



\maketitle

\begin{abstract}

This work addresses the landing problem of an aerial vehicle, exemplified by a simple quadrotor, on a moving platform using image-based visual servo control. First, the mathematical model of the quadrotor aircraft is introduced, followed by the design of the inner-loop control. At the second stage, the image features on the textured target plane are exploited to derive a vision-based control law. The image of the spherical centroid of a set of landmarks present in the landing target is used as a position measurement, whereas the translational optical flow is used as velocity measurement. The kinematics of the vision-based system is expressed in terms of the observable features, and the proposed control law guarantees convergence without estimating the unknown distance between the vision system and the target, which is also guaranteed to remain strictly positive, avoiding undesired collisions. The performance of the proposed control law is evaluated in MATLAB and 3-D simulation software Gazebo. Simulation results for a quadrotor UAV are provided for different velocity profiles of the moving target, showcasing the robustness of the proposed controller.
\end{abstract}

\section{Introduction}
In recent years, Unmanned
Aerial Vehicles (UAV) cooperative surveillance projects with marine vehicles have been developed for
its potential applications, with the aerial vehicle providing a bird’s eye view of an area of interest that
can complement and guide the marine vehicle in its operations. These operations frequently extend over large periods of time, which turn out to be a bottleneck for UAVs, as most of these aerial vehicles lack the flight endurance and range for sustained flight. From this, there have been modern advances on developing hybrid-fueled UAVs \cite{DeWagter:2021:neder} to tackle the flight endurance problem, however they bring new aerodynamical challenges to the general multi-rotor UAV model, considering the addition of fuel cells and large hydrogen storage equipment. A possible scheme to mitigate this general UAV limitation is to reduce the flight time and allow the aerial vehicle to take-off and land on a moving landing platform as required by the surveillance mission.

Accurately measuring an aerial vehicle's position within its local environment poses a significant challenge when designing motion control systems. While outdoor UAVs often rely on global positioning systems (GPS), it falls short in providing comprehensive positioning information, ignoring local terrain features and proving to be unreliable in cluttered environments. To overcome these limitations, integrating a vision system with an Inertial Measurement Unit (IMU), commonly incorporated in UAVs for attitude and angular velocity estimation, offer a practical solution for advancing autonomous navigation across diverse environmental conditions.
\subsection{Related Work}

    The integration of a vision system directly in the formulation
    of the control laws without attempting to estimate the position
    and velocity is denominated image-based visual servo (IBVS)
    control \cite{Chaumette:2006:IBVS,Mahony:2008:IBVS}. This field of research has been extensively developed over the last years with the use of optical flow as velocity measurement for hovering and vertical landing operations \cite{Herisse:2008:opticalflow}, as well as extended to landing on moving targets in \cite{Cho:IBVS_FF:2022,Daewon:2012:IBVS, Mahony:2012:Landing, Serra:2014:opticalflow_IEEE, Serra:2016:Primeiroqueli}. In particular, in \cite{Cho:IBVS_FF:2022}, the moving ship's velocity is estimated and used as a feed-forward term in the IBVS control and adaptive methods are applied to achieve a more robust vision-based landing procedure. In \cite{Mahony:2012:Landing}, a measurement of the average optical flow in spherical coordinates is exploited to enable hover and landing control on the moving target platform. Furthermore, in \cite{Serra:2014:opticalflow_IEEE, Serra:2016:Primeiroqueli}, a dynamic IBVS-IMU-assisted control method is proven to completely reject bounded and possibly time-varying disturbances. This robustness property of the proposed control law is highly relevant in the scenario of landing an UAV on a moving ship in the presence of maritime disturbances.

\begin{figure}[h]
    \centering
    \includegraphics[trim={0 10.5cm 0 0}, clip, scale=0.07]{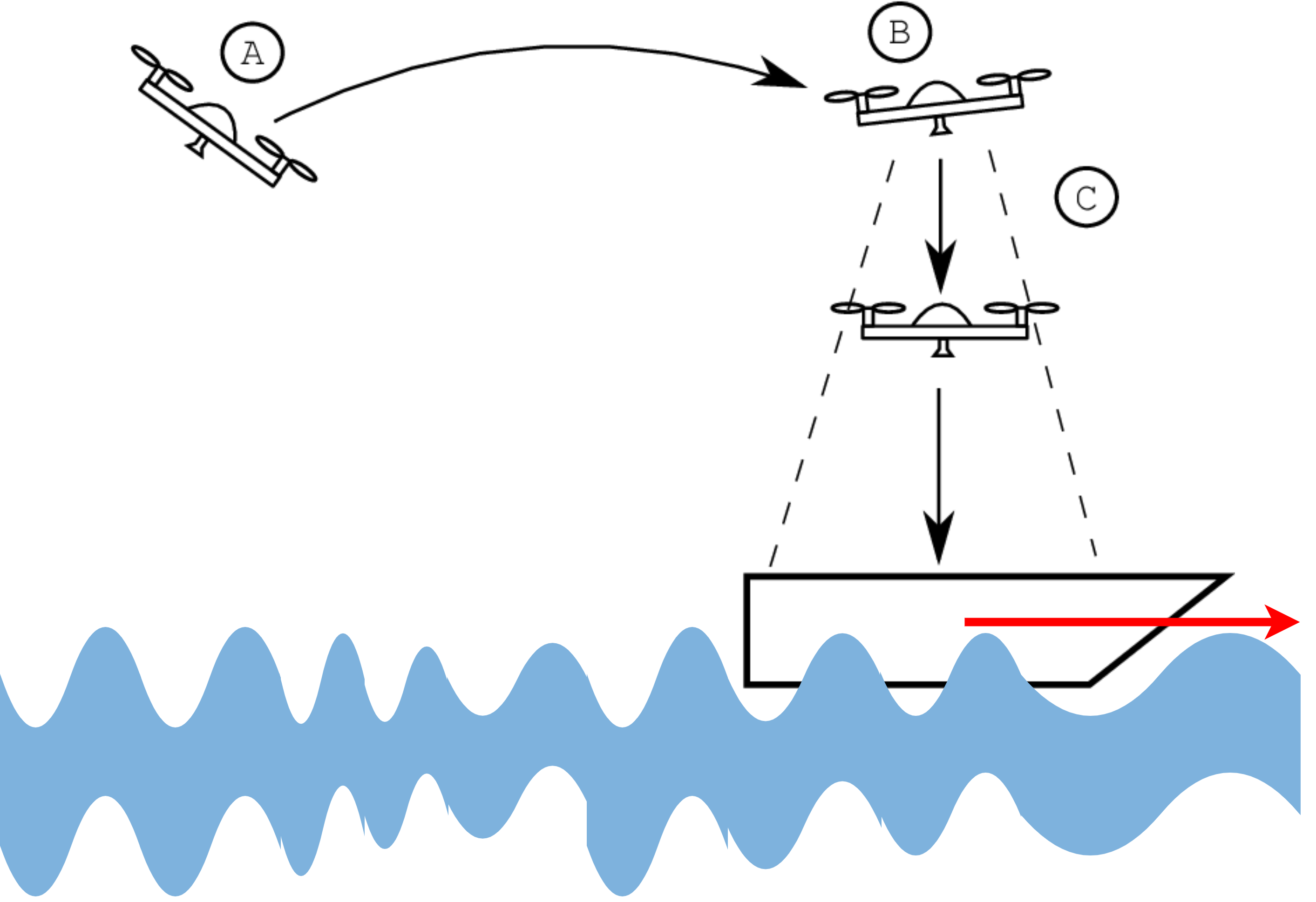}
    \caption{Landing manoeuvre of an UAV on a moving target}
    \label{landing_operation}
\end{figure}

Motivated by all of those factors, the aim of the present work is to develop a set of tools that allow a quadrotor UAV to perform landing missions using only inertial and image data, where the main goal is to ensure a smooth landing on a moving platform, avoiding undesired collisions, according to Figure \ref{landing_operation}.

\section{UAV Modeling}
\begin{figure}[h]
    \centering
    \includegraphics[scale=0.15]{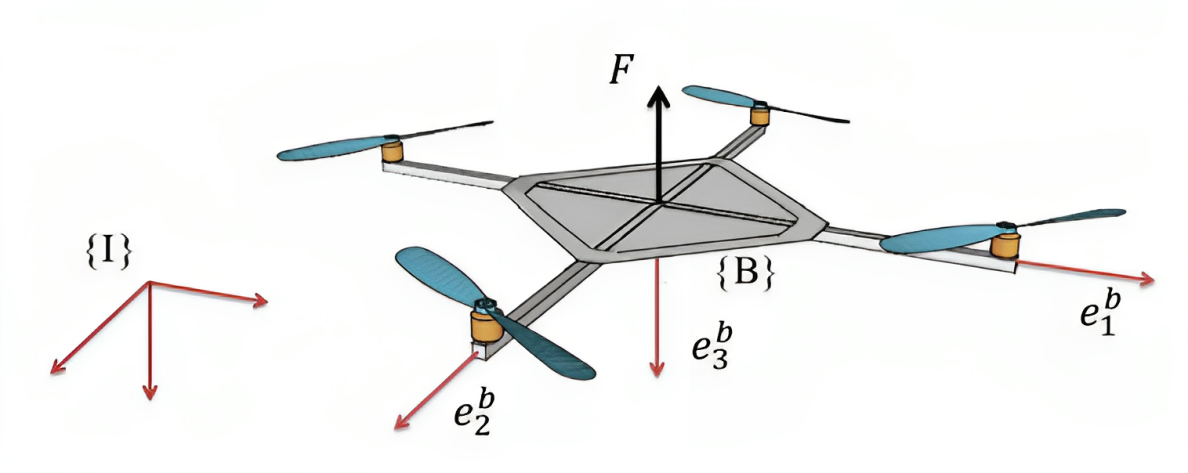}
    \caption{Reference frames and force for schematic representation of a quadrotor}
    \label{fig:quadrotor_model}
\end{figure}

To describe the motion of the UAV model, two reference frames are introduced: the inertial reference frame $\{\mathcal{I}\}$ fixed to the earth surface in North-East-Down (NED) convention and the body reference frame $\{\textit{B}\}$ attached to the aerial vehicle's center of mass. Consider a quadrotor vehicle consisting of a rigid cross-frame equipped with four rotors (Figure \ref{fig:quadrotor_model}). 

The system can be modeled as a rigid body of mass $m$ and tensor of inertia $I$, subject to external forces and moments generated by the rotors and due to gravity. The kinematics of the quadrotor can be written as
\begin{equation}
    \begin{bmatrix}
             \dot{\xi} \\
             \dot{E}
    \end{bmatrix} = \begin{bmatrix}
        R & 0_{3\times3} \\
        0_{3\times3} & Q
    \end{bmatrix}\begin{bmatrix}
        V \\
        \Omega
    \end{bmatrix}
    \label{sect2:eq_kinematics}
\end{equation}
where with respect to the inertial reference frame {$\mathcal{I}$}, $\xi=[x\ y\ z]^T$ denotes the position of the origin of \{\textit{$B$}\}, $E$ = $[\phi \ \theta \ \psi]^T$  the orientation of \{\textit{$B$}\} expressed in $ZYX$ Euler angles (roll, pitch, yaw), $V$ the translational velocity of the origin of \{\textit{$B$}\}, $\Omega=[p\ q\ r]^T$ the angular velocity of the origin of \{\textit{$B$}\}, $R\in SO(3)$ the rotation matrix of the body-fixed frame \{\textit{$B$}\} and $Q$ the angular velocity transformation matrix from \{\textit{$B$}\} to \{$\mathcal{I}$\}. The dynamics equations can be given by
\begin{align}
    m\dot{V} &= -m \Omega_\times V + F + R^T\mu_w, \label{sect2:eq_final_trans_dynamics} \\
    I\dot{\Omega} &= -\Omega_\times I \Omega + \Gamma, \label{sect2:eq_final_rotational_dynamics}
\end{align}
where $\dot{V}$ denotes the translational acceleration, $F$ the force applied to the aerial vehicle, $I$ is the inertia matrix, $\Gamma \in \mathbb{R}^3$ is the torque applied to the aerial vehicle, and $\Omega_\times$ the skew symmetric matrix associated with the vector product $\Omega_\times = \Omega \times x = [p\ q\ r]^T \times x$, for any $x \in\mathbb{R}^3$. The term $\mu_w(t)$ is expressed in the inertial frame \{$\mathcal{I}$\} and accounts for disturbances such as modeling errors, aerodynamical effects, wind disturbances, and others. 

The force vector $F$ combines thrust, gravity and aerodynamical components and can be modeled as
\begin{equation}
    F = -F_T+mgR^Te_3+F_{aero},
    \label{sect2:eq_translational_force}
\end{equation}
where $g$ is the gravitational acceleration, $F_T\in \mathbb{R}^3$ is the thrust control input, and $e_3=[0\ 0\ 1]^T$. $F_{aero}$ denotes the aerodynamic disturbances and tend to be second order compared to the dominant thrust control \cite{Mahony:2008:IBVS}, and modeling such terms is highly complex and would significantly draw back attention from the proposed IBVS control design, as such, the aerodynamic term is ignored in the control design and regarded as part of the load disturbance $\mu_w$. 

\section{Inner-Loop Design}
\begin{figure}[h]
    \centering
    \includegraphics[scale=0.4]{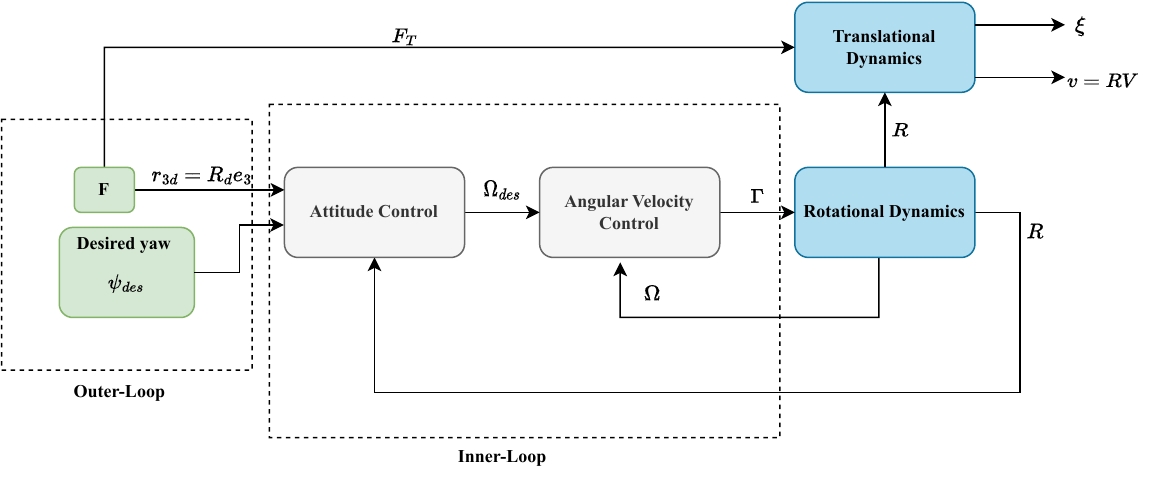}
    \caption{Hierarchical control design strategy}
    \label{sect3:fig:control_strategy}
\end{figure}

Given the quadrotor thrust configuration, it is common to design an inner-loop controller whose task is to follow attitude reference commands. This method is based on the assumption that the controllers will be tuned such that the closed-loop attitude dynamics converges faster than the closed-loop translational dynamics, by using, for example, a 'high-gain' attitude controller. The inner loop continuously regulates the torque inputs to track a desired orientation $R_d(t)$, and the desired orientation $R_d(t)$ is chosen to provide a thrust input with the adequate direction for translational dynamics \eqref{sect2:eq_final_trans_dynamics}, which determines $r_{3d}(t):=R_d(t)e_3$ and a desired yaw angle. More precisely, for the translational dynamics of the UAV quadrotor \eqref{sect2:eq_final_trans_dynamics}, the force $F$ \eqref{sect2:eq_translational_force} is used as the control input by means of its thrust direction and magnitude. Given a desired force $F_d$, \eqref{sect2:eq_translational_force} can be solved for $F_T$ and $r_{3d}$. With a suitable high-gain inner-loop control that ensures that the attitude dynamics is fast enough to assume that $R=R_d$, it is possible to consider that the force input $F$ converges fast enough to the desired force $F_d$, $F \approx F_d$ \cite{Ghanadan:IEEE:1994:Smallgains, Sepulchre:Book:1996:Smallgains}. Figure \ref{sect3:fig:control_strategy} presents the hierarchical control architecture adopted. 
\begin{problem}
Consider a quadrotor described by \eqref{sect2:eq_final_trans_dynamics}, \eqref{sect2:eq_final_rotational_dynamics} and let $E_d = [\phi_d\ \theta_d\ \psi_d]^T$ denote the desired roll, pitch, and yaw angles, respectively. Linearize vehicle dynamics and design a linear controller for torque control $\Gamma \in \mathbb{R}^3$ such that $E$ converges to the desired angle references $E_d$.
\end{problem}
\begin{assumption}
    The quadrotor UAV's inertia matrix $I$ is diagonal, i.e, the rotational motion around one axis does not affect the rotational motion around another axis.
    \label{sect3:assumption:inertia_diagonal}
\end{assumption}
\begin{assumption}
The quadrotor UAV is assumed to be operating around its equilibrium point (hover state), i.e. $\theta \approx 0$ and $\phi \approx 0$.
\label{sect3:assumption:_hover_state_eq_point}
\end{assumption}
Considering assumptions \ref{sect3:assumption:inertia_diagonal}, \ref{sect3:assumption:_hover_state_eq_point}, it is possible to decouple the quadrotor's rotational motion around the three axis, allowing separate control of the roll, pitch and yaw motions \eqref{sect2:eq_final_rotational_dynamics}, and near hover state, the quadrotor's linearized angular acceleration can be written as
\begin{equation}
    \ddot{E}=I^{-1}\Gamma,
\end{equation}
such that $\Gamma=I\cdot u_m$, where $u_m$ represents the control input. The proposed Proportional Derivative (PD) inner-loop control law 
\begin{equation}
    u_m(t)=K_pe(t)+K_d\dot{e}(t) \text{ with } K_p, K_d \succeq 0, 
    \label{sect3:eq_control_input_moment}
\end{equation}
where $e(t) = E_d(t)-E(t)$ and $\dot{e}(t) = \dot{E}_d(t)-\dot{E}(t)$. 
\subsection{Generating angle references from acceleration commands}
When designing an outer-loop for an UAV quadrotor, it is common to adopt a simple double integrator system model since the aerial vehicle is free to move in 3-D space and can be controlled at the force level. In this system, the position error is driven to zero by controlling the acceleration of the vehicle, whereas $\psi_{des}$ is left as a free variable controlled by the outer-loop controller. Consider the double integrator model of the aerial vehicle given by
\begin{equation}
    \ddot{p}:=\ddot{\xi}=u_f=\frac{F}{m}=-\frac{F_{thrust}}{m}Re_3+ge_3,
    \label{sect3:eq_double_integrator_model}
\end{equation}
where $u_f$ is the output of the outer-loop controller and $g$ is the gravitational acceleration. It is necessary to develop a sub-system capable of computing the total thrust $F_{thrust}$ and the attitude associated with the matrix $R$ from the desired input of the outer-loop. \eqref{sect3:eq_double_integrator_model} can be expanded to:
\begin{equation}
    u_f =\frac{F}{m}= -\frac{1}{m}R_z(\psi_{des})[R_y(\theta)R_x(\phi)F_{thrust}e_3]+ge_3.
    \label{sect3:eq_control_input_force}
\end{equation}
Furthermore, consider $u^\star$ to be given by
\begin{equation}
    u^\star :=R_y(\theta)R_x(\phi)F_{thrust}e_3.
    \label{sect3:eq_control_input_star}
\end{equation}
Using \eqref{sect3:eq_control_input_star} in \eqref{sect3:eq_control_input_force}, 
\begin{equation}
    u^\star = -mR_z^T(\psi_{des})(u_f-ge_3).
    \label{sect3:eq_control_input_star_final}
\end{equation}
From \eqref{sect3:eq_control_input_star_final}, it is possible to compute $u^\star$ from the desired output of the outer-loop controller. Considering that the total thrust is given by $F_{thrust}:=\norm{u^\star}$, the desired angle references can be obtained from 
\begin{equation}
    \phi_{des}=asin(\frac{-u^\star_2}{F_{thrust}}),
\end{equation}
\begin{equation}
    \theta_{des} = atan(\frac{u^\star_1}{u^\star_3}), u^\star_3 \neq 0,
\end{equation}
where $asin(.)$ denotes the inverse trigonometric function $sin(.)$, $atan(.)$ denotes the inverse trigonometric function $tan(.)$, and $u^\star=[u^\star_1, u^\star_2, u^\star_3]^T$ denotes the virtual control input. 
With the adopted double integrator system model, angle reference commands are calculated from the outer loop controller output and fed into the hierarchical control loop (Figure \ref{sect3:fig:control_strategy}).
\section{Image Features}
In this section, we introduce the pinhole camera and the spherical camera model adopted to describe image formation. The optical flow equations are provided and the visual features used for control purposes are derived. 

\begin{figure}
    \centering
    \includegraphics[scale=0.15]{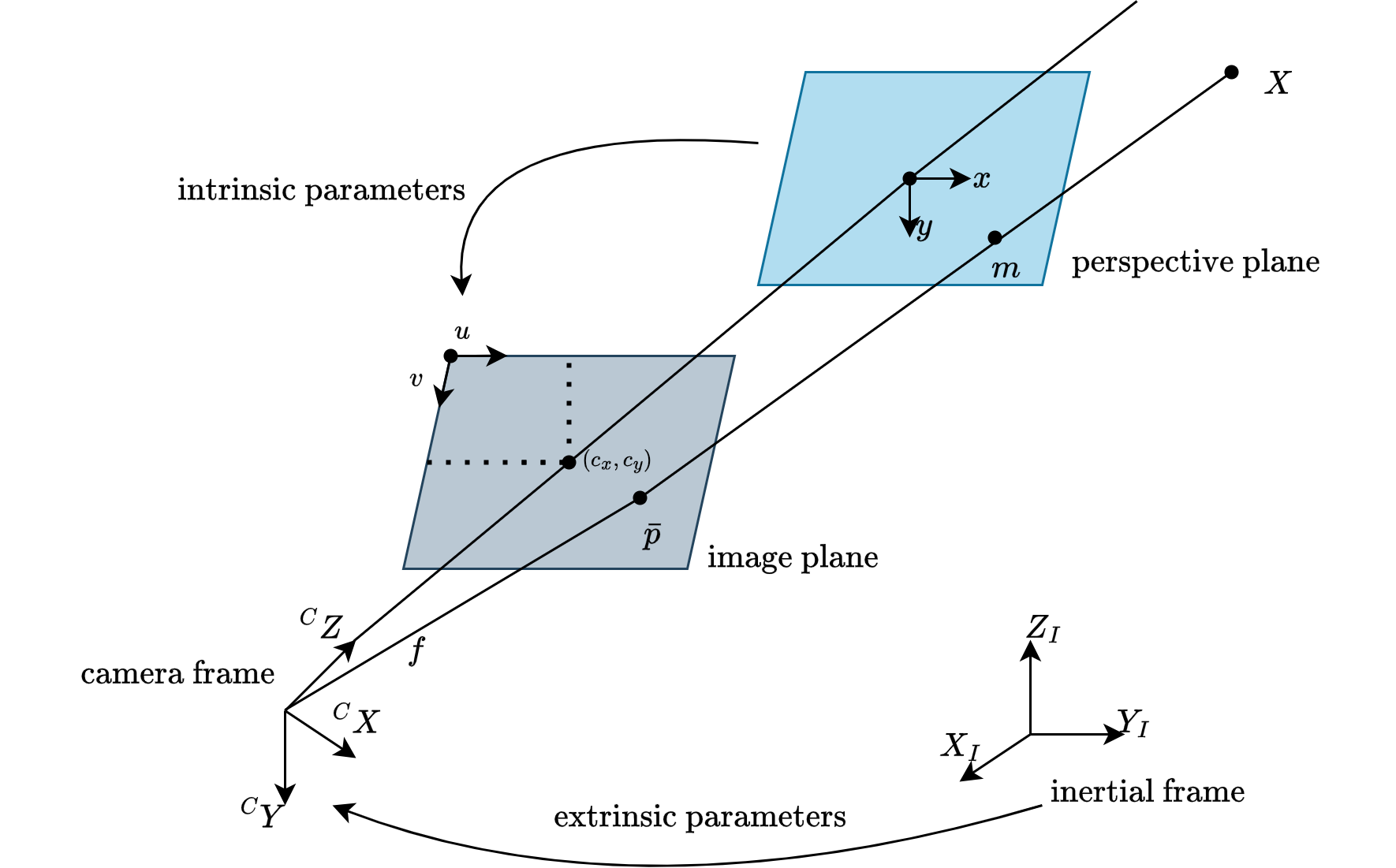}
    \caption{Pinhole camera model}
    \label{sect4:fig:camera_model}
\end{figure}

Consider a perspective camera model characterized by a set of intrinsic parameters that define the mapping from camera coordinates to the image plane, and extrinsic parameters that establish the relationship between coordinates in the inertial frame \{$\mathcal{I}$\} and the camera frame \{$C$\}. Figure \ref{sect4:fig:camera_model} presents an illustration of the pinhole camera model.
A 3-D point ${}^{I}P=[X_I\ Y_I\ Z_I ]^T$ in the inertial frame can be expressed in homogeneous coordinates and projected to an image by
\begin{equation}
    \lambda \begin{bmatrix}
        x \\
        y \\
        1
    \end{bmatrix}= \underbrace{ \begin{bmatrix}
        fs_x & 0 & c_x \\
        0 & fs_y & c_y \\
        0 & 0 & 1
    \end{bmatrix} }_{K}\begin{bmatrix}
        1 & 0 & 0 & 0 \\
        0 & 1 & 0 & 0 \\
        0 & 0 & 1 & 0 
    \end{bmatrix}\begin{bmatrix}
        {}^C_{I} R & {}^C_{I} T \\
        0_{1\times 3} & 1
    \end{bmatrix}\begin{bmatrix}
        X_I \\
        Y_I \\
        Z_I \\ 
        1
    \end{bmatrix},
    \label{ch5:complete_equation_of_perspective}
\end{equation}
where $\lambda=\frac{1}{Z}$ is a scale factor encoding depth information, $x$ and $y$ denote the coordinates of the point ${}^{I}P$ in the image frame, $K\in \R^{3\times 3}$ denotes the matrix of intrinsic parameters and ${}^C_{I} [R|T]$ represents the rigid-body transformation (rotation and translation) of coordinates expressed in the inertial frame \{$\mathcal{I}$\} to the camera frame \{$C$\}. In the camera frame \{$C$\}, the third component of an image point is always equal to the focal length $z = f$. Thus, it is common to represent the image point as $q = [x\ y]^T$ instead of $q =[x \ y \ f]^T$ where
\begin{align*}
    x &= f\frac{X}{Z} \\
    y &= f\frac{Y}{Z},
\end{align*}
or expressed in matrix notation as:
\begin{equation}
    \begin{bmatrix}
        q \\ 
        1
    \end{bmatrix}=\lambda \begin{bmatrix}
        f & 0 & 0 \\
        0 & f & 0 \\ 
        0 & 0 & 1 
    \end{bmatrix}{}^CP,
\end{equation}
where ${}^CP=[X_C\ Y_C\ Z_C]^T$ is a 3-D point expressed in the camera frame. The planar image point $q=[x\ y]^T$ can be converted to pixel coordinates $\bar{p} \in \R^2$:
\begin{equation}
    \begin{bmatrix}
        \bar{p} \\
        1
    \end{bmatrix}=A\begin{bmatrix}
        q \\
        1
    \end{bmatrix}
\end{equation}
where 
\begin{equation*}  
    A = \begin{bmatrix}
        s_x & 0 & c_x \\
        0 & s_y & c_y \\
        0 & 0 & 1
    \end{bmatrix},
\end{equation*}
and ${}^CP \in \R^3$ can be transformed into 2-D pixel coordinates from 
\begin{equation}
    \begin{bmatrix}
        \bar{p} \\
        1
    \end{bmatrix}=\lambda K {}^CP.
    \label{ch5:eq:pixel_coordinates}
\end{equation}

\subsection{Spherical Camera Geometry}
\begin{figure}[H]
    \centering
    \includegraphics[scale=0.5]{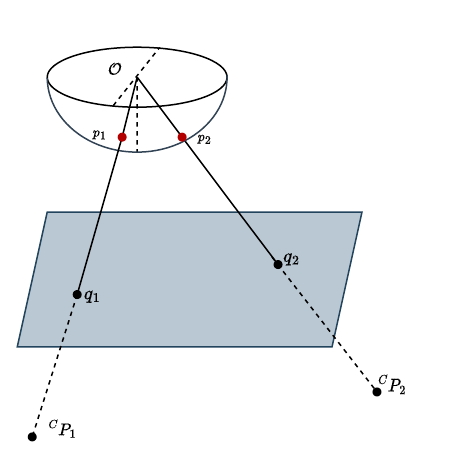}
    \caption{Spherical image projection}
    \label{ch5:fig:spherical_image_projection}
\end{figure}

Consider the point ${}^CP=[X_C\ Y_C\ Z_C]^T\in \R^3$ expressed in \{$C$\} and let $\bar{p} \in \R^2$ denote its corresponding image point in pixel coordinates. From \eqref{ch5:eq:pixel_coordinates}, the planar image point coordinates $q \in \R^2$ is given by
\begin{equation}
        \begin{bmatrix}
        q \\ 
        1 
    \end{bmatrix}=K^{-1}\begin{bmatrix}
        \bar{p} \\
        1
    \end{bmatrix}=\frac{{}^CP}{Z}.
    \label{ch5:eq:5.1.1_planar_imagepoint}
\end{equation}
To obtain the spherical projection of the image point $q$, we simply normalize $q$
\begin{equation}
    p=\frac{[
        q\ 1
    ]^T}{\norm{[
        q\ 1
    ]^T}}.
\end{equation}
From \eqref{ch5:eq:5.1.1_planar_imagepoint}, ${}^CP = Z[
    q\ 1
]^T$ and $\norm{{}^CP}=Z\norm{[
    q\ 1
]^T}$.
Thus, the spherical image point $p$ can be directly obtained from the 3-D point ${}^C P$
\begin{equation}
    p = \frac{Z[q\ 1]^T}{Z\norm{[q\ 1]^T}}=\frac{{}^CP}{\norm{{}^CP}}.
    \label{eq: spherical_projection_from_3d_point}
\end{equation}
Take, for instance, the example in Figure \ref{ch5:fig:spherical_image_projection} where two planar image points, $q_1$ and $q_2$, are projected onto the unit sphere, denoted as $p_1$ and $p_2$. 
\subsection{Optical Flow}
The optical flow (OF) of an image is a 2-D velocity vector field which describes the motion of each pixel between frames, caused either by the movement of the viewer, i.e. the camera, or the detected objects in the environment. The classical calculation of OF works on multiple assumptions. 
\begin{assumption}
    The observed brightness of any object point is constant over a small time step, $B(x(t), y(t), t)=\text{constant}$. Hence $\frac{dB}{dt}=0$ for a small time step. 
\end{assumption}
By assuming brightness constancy and small motion, the OF is calculated from the changes of the image brightness $B(x, y, t)$:
\begin{equation}
    \frac{dB}{dt}=\begin{bmatrix}
        \frac{\partial B}{\partial x} & \frac{\partial B}{\partial y}
    \end{bmatrix}\cdot \begin{bmatrix}
        p_x \\
        p_y
    \end{bmatrix} +\frac{\partial B}{\partial t} = 0,
    \label{ch5:equation_image_brightness}
\end{equation}
where $p_x=\frac{dx}{dt}$ and $p_y=\frac{dy}{dt}$ are the optical flow velocities. The second assumption imposes a relationship between neighboring pixels, in which they have similar motion.  
\begin{assumption}
    The local optical flow is constant over an ($n\times n\times  \Delta t$) neighborhood of pixels around the desired pixel $(x, y, t)$, where $n$ is the size of the patch around the desired pixel (local smoothness of brightness). 
\end{assumption}
The \textit{Optical Flow} equation is defined as 
\begin{equation}
    f_xp_x + f_yp_y + f_t = 0, 
    \label{eq:optical_flow_equation}
\end{equation}
where $f_x = \frac{\partial B}{\partial x}$,  $f_y = \frac{\partial B}{\partial y}$ are the image gradients, 
$f_t = \frac{\partial B}{\partial t}$ the gradient along the time, and $p_x = \frac{dx}{dt}$ and $p_y=\frac{dy}{dt}$ being the unknown 2-D optical flow. 
\subsubsection{Lucas-Kanade Algorithm}
The Lucas-Kanade method takes a patch $W$ of size $(n\times n),\ n>1$ around the extracted point of the image and assumes that all the $n^2$ points have the same motion. Thus, the problem in \eqref{eq:optical_flow_equation} turns into solving $n^2$ equations with two unknown variables $(p_x,p_y)$, which is an over-determined system. For all points $(k,l)\in W$:
\begin{equation}
\underbrace{
    \begin{bmatrix}
        f_x(1,1) & f_y(1,1) \\
        f_x(k,l) & f_y(k,l) \\
        \vdots & \vdots \\
        f_x(n, n) & f_y(n, n)
    \end{bmatrix}
    }_{A}
    \underbrace{
    \begin{bmatrix}
        p_x \\
        p_y
    \end{bmatrix}
    }_{u}=\underbrace{\begin{bmatrix}
        f_t(1,1) \\
        f_t(k,l) \\
        \vdots \\
        f_t(n,n)
    \end{bmatrix}.
    }_{B}
    \label{ch5:eq_lucas_kanade}
\end{equation}
\eqref{ch5:eq_lucas_kanade} can be described in the form of a linear system:
\begin{align*} 
Au&=B \\ 
A^TAu&=A^TB,
\end{align*}
and a solution can be obtained using the least-square method,
\begin{equation}
\underbrace{\begin{bmatrix}
        \sum_W f_{x}^2 & \sum_W f_{x}f_{y} \\
        \sum_W f_{x}f_{y} & \sum_W f_{y}^2
    \end{bmatrix}}_{A^TA}
    \underbrace{\begin{bmatrix}
        p_x \\
        p_y
    \end{bmatrix}}_{u}=
    \underbrace{\begin{bmatrix}
        -\sum_W f_{x}f_{t} \\
         -\sum_W f_{y}f_{t}
    \end{bmatrix}}_{A^TB},
\end{equation}
\begin{equation}
    u=(A^TA)^{-1}A^T.\label{ch5:eq_least_square}
\end{equation}
In practice, the optical flow estimation is unreliable in smooth regions where the equations for all pixels in the chosen window are approximately the same, and edges where the image gradient is prominent in one direction. Resorting to the computer vision library OpenCV \cite{OPENCV:website} Lucas-Kanade pyramidal implementation or optical flow sensors \cite{optical_sensors}, we can specify a set of feature points in the image and calculate the 2-D optical flow $\dot{\bar{p}}=(p_x, p_y)$, to be used for control purposes. 
\subsection{Spherical Image Kinematics}
\begin{figure}
    \centering
    \includegraphics[scale=0.5]{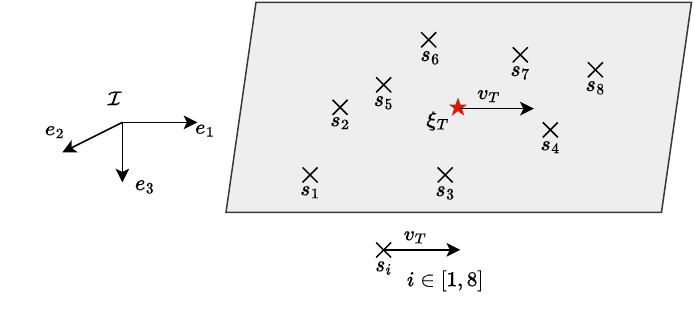}
    \caption{Landmarks on the target plane}
    \label{ch5:fig:landmarks}
\end{figure}

Let $s_i \in \mathbb{R}^T$ denote the position of the $i^{\text{th}}$ landmark of the target plane relative to the reference inertial frame expressed in \{$\mathcal{I}$\}, $\xi_T = \frac{1}{n}\sum_{i=1}^n s_i$ the centroid of the landmarks and $\xi$ the position of the quadrotor measured in \{$\mathcal{I}$\}. The position of the $i^{\text{th}}$ landmark of the target relative to the body-fixed frame \{\textit{$B$}\} is
\begin{equation}
    P_i = R^T(s_i-\xi) \in \mathbb{R}^3.
    \label{eq:P_i}
\end{equation}
The centroid of the landmarks relative to the body-fixed frame is given by
\begin{equation}
    \mathcal{C} = \frac{1}{n}\sum_{i=1}^n P_i
    \label{eq: centroid_landmarks}
\end{equation}
where $n\geq 1$ is the number of landmarks, according to Figure \ref{ch5:fig:landmarks}.

An observed landmark from the target plane inherits kinematics from the motion of both the camera observer and the landing platform. From \eqref{eq:P_i} and (\eqref{sect2:eq_kinematics} and \eqref{sect2:eq_final_trans_dynamics}), describing the dynamics of a quadrotor aerial vehicle, the kinematics of $P_i$ can be written as 
\begin{equation}
    \dot{P_i} = -\Omega_\times P_i + (V_T-V)
    \label{eq:kinematics_Pi}
\end{equation}
where $V_T=R^Tv_T$ is the translational velocity of the target plane and $V$ is the linear velocity of the aerial vehicle, both expressed in \{\textit{$B$}\}. With this, the kinematics of the spherical image point $p_i$ can be derived from \eqref{eq: spherical_projection_from_3d_point} and \eqref{eq:kinematics_Pi}:
\begin{equation}
    \dot{p_i} = -\Omega_\times p_i -\pi_{p_i}\frac{1}{\norm{P_i}}(V-V_T).
    \label{eq:kinematics_pi}
\end{equation}
where $\pi_x$ denotes the orthogonal projection operator onto the tangent space to the sphere at $x$
\begin{equation*}
\pi_x = I_3 -xx^T
\end{equation*}
such that for any point $g\in \mathbb{R}^3$, $\pi_{p_i}g$ generates the projection of $g$ onto the plane orthogonal to $p_i$. Moreover, let $d$ denote the height of the vehicle above an observed landmark of the landing plane (which is assumed to be horizontal)
\begin{equation}
    d:=\eta^TP_i.
    \label{ch5:eq:height_1}
\end{equation}
The unit normal to the target plane expressed in the body-fixed frame is given by $\eta$. From \eqref{eq: spherical_projection_from_3d_point}, the height $d$ can be formulated as
\begin{equation}
    d(t) = \eta^Tp_i\norm{P_i} = cos(\theta_i)\norm{P_i},
    \label{eq: height_2}
\end{equation}
where $\theta_i$ is the angle between the normal direction $\eta$ to the target plane and the observed image point $p_i$ from the target plane. Substituting this relationship into \eqref{eq:kinematics_pi} yields
\begin{equation}
    \dot{p_i} = -\Omega_\times p_i - \frac{cos(\theta_i)}{d(t)}\pi_{p_i}(V-V_T).
    \label{eq: kinematics_pi_2}
\end{equation}
\subsection{Visual Velocity Measurement --- Translational Optical Flow}\label{section:w_computation}
\begin{figure}
    \centering
    \includegraphics[scale = 0.3]{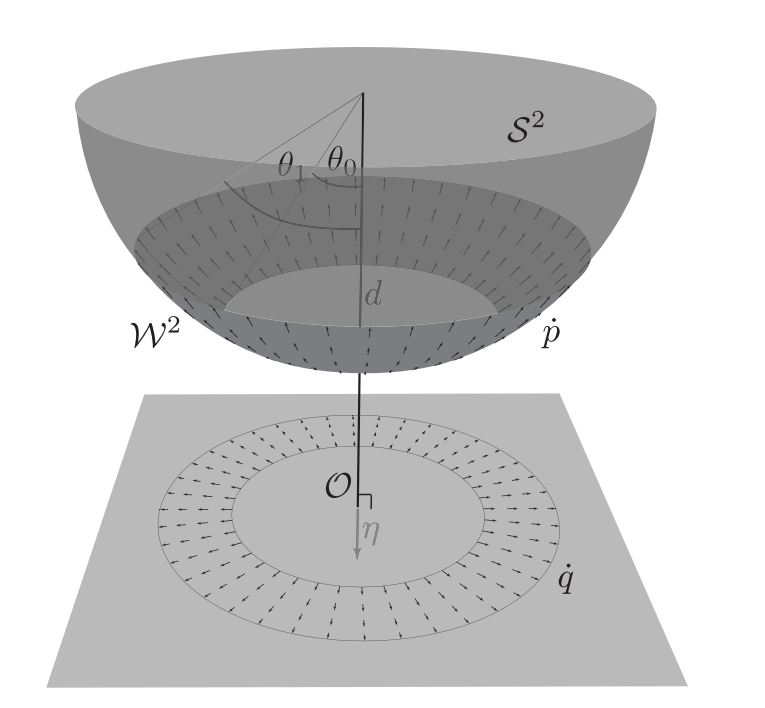}
    \caption[Translational optical flow computation]{Translational optical flow computation (Image from \cite{Serra:2016:Primeiroqueli})}
    \label{fig: translational_optical_flow}
\end{figure}

The visual velocity measurement that is used in the control law is the translational optical flow given by
\begin{equation}
    W(t) = \frac{V(t)-V_T(t)}{d(t)}
    \label{eq: translational_optical_flow}
\end{equation}
and \eqref{eq: kinematics_pi_2} can be revised to
\begin{equation}
    \dot{p_i} = -\Omega_\times p_i - cos(\theta_i)\pi_{p_i}W.
\end{equation}
An effective measurement of $W$ is obtained from the integral of the observed optical flow over the solid area $\mathcal{W}^2$ of the lower hemisphere of the sphere $\mathcal{S}^2$ corresponding to the visual image of a region of the target plane, with central point $\mathcal{O}$, as depicted in Figure \ref{fig: translational_optical_flow}.
The observed optical flow over the solid angle $\mathcal{W}^2$ can then be defined by 
\begin{equation}
\begin{split}
    \phi_w & = \int \int_{\mathcal{W}^2} \dot{p}dp \\
    & = \int_0^{2\pi} \int_{\theta_0}^{\theta_1}(-\Omega_\times p - cos(\theta)\pi_{p}W)sin(\theta)d\theta d\alpha.
\end{split}
\label{eq:integral_dot_p}
\end{equation}
The factor $sin(\theta)$ accounts for the geometry of the spherical coordinate system, the angles parameters $0\leq\theta_0<\theta_1$ and $\theta_1\leq \frac{\pi}{2}$ define the area of integration and are here for optimizing the practical implementation of the algorithm. Increasing the angle $\theta_0$ results in ignoring image points close to the central point $\mathcal{O}$ which do not carry much information about vertical translation since their velocities projected onto the sphere are approximately 2-D vectors and will also reduce the computational cost of solving the integral. The value of $\theta_1$ is limited by the field of view (FOV) of the camera considering that it is necessary for the points to be within the FOV to accurately compute their local optical flow. A larger integral area, in principle, reduces the effect of noise on the computation.

Lengthy but straightforward calculations show that the optical flow on the solid angle $\mathcal{W}^2$ is given by
\begin{equation}
    \phi_w = -\frac{\pi}{2}(cos(2\theta_0)-cos(2\theta_1))\Omega_\times \eta + R^T\Lambda RW
    \label{eq: phi_w_last}
\end{equation}
where the matrix $\Lambda$ is a constant diagonal matrix depending on the parameters of the solid angle $W^2$.
Solving for $W$ \eqref{eq: phi_w_last} yields
\begin{equation}
    W = -R^T\Lambda^{-1}R(\phi_w+\frac{\pi}{2}(cos(2\theta_0)-cos(2\theta_1))\Omega_\times \eta).
    \label{eq:derotation}
\end{equation}

In practice, the measurement of the optical flow under spherical projection $\dot{p}$ constitutes a fundamental aspect of the proposed algorithm. The Lucas-Kanade algorithm allows a fast and precise calculation of the 2-D optical flow field of a specific set of feature points resulting in a set of velocities. However, note that the effectiveness of the approach also depends on the specifications of the camera, namely its focal length, and sufficient texture of the target plane to obtain a fairly dense measurement of optical flow $\dot{\Bar{p}}=(p_x, p_y)$. In optical flow applications, it's crucial for each initial image point to be easily distinguishable from others in the image, and the mapping of its motion to a final point should be done independently. By ensuring accurate tracking of individual image points, the algorithm provides more accurate optical flow measurements. 
\subsection{Visual Position Measurement --- Centroid Vector }
The visual feature used in the control law, which encodes information about the position of the vehicle relative to the target plane, is the centroid vector given by 
\begin{equation}
    q:=\frac{1}{n}\sum_{i=1}^n \frac{p_i}{\eta^T p_i} \in \mathbb{R}^3.
    \label{eq: centroid_vector_q}
\end{equation}
Substituting \eqref{eq: centroid_landmarks} and \eqref{eq: height_2} into \eqref{eq: centroid_vector_q} results in
\begin{equation*}
    q = \frac{\mathcal{C}}{d}.
\end{equation*}
The derivative of the centroid position of the landmarks relative to \{\textit{$B$}\} \eqref{eq: centroid_landmarks} is given by 
\begin{equation}
    \dot{\mathcal{C}} = -\Omega_\times\mathcal{C}-(V-V_T).
    \label{ch5:eq:time_derivative_centroid_landmarks}
\end{equation}

The main objective is to design a feedback controller resorting only to the derived visual measurements that can ensure automatic landing of the aerial vehicle. In this sense, the control goal for landing purposes is driving $\mathcal{C}$ to zero and ensuring that the vehicle's velocity $V$ converges to the target plane's velocity $V_T$, i.e. land on the centroid of the landmarks of the target plane without undesired physical collisions.

\section{Vision-Based Control}
In this section, we consider the problem of designing a feedback control law for the landing operation using the visual position measurement -- centroid vector in the camera frame $q$, the visual velocity measurement -- translational optical flow $W$, and the normal vector to the target $\eta$.
\begin{problem}
    Consider a quadrotor UAV with kinematics and dynamics described by \eqref{sect2:eq_kinematics}, \eqref{sect2:eq_final_trans_dynamics}, \eqref{sect2:eq_final_rotational_dynamics} and is equipped with a control structure that given a desired bounded acceleration $u_f^d\in \mathbb{R}^3$, computes an attitude and thrust reference for the inner-loop system \eqref{sect3:eq_double_integrator_model}. Design a vision-based control law for the quadrotor acceleration command such that:
    \begin{itemize}
        \item the aerial vehicle lands on the centroid of the landmarks, i.e. $\mathcal{C}$ converges to zero;
        \item the velocity of the quadrotor $V$ converges to the velocity of the target plane $V_T$, i.e. $|V-V_T|\rightarrow 0$ as $t\rightarrow \infty$; and
        \item the height $d$ between the aerial vehicle and the target remains strictly positive at all times, and $d=0$ is reached with zero relative velocity.
    \end{itemize} 
\label{sect5:problem:image_dynamics_system}
\end{problem}
Defining the relative velocity:
\begin{equation}
    \Tilde{V}:=V-V_T,
\end{equation}
and considering the equations introduced in \eqref{sect2:eq_final_trans_dynamics} and 
the image kinematics and dynamics can be described as
\begin{equation}
    \begin{cases}
    \dot{\mathcal{C}} &= -\Omega_\times\mathcal{C}-\Tilde{V} \\
    \dot{\Tilde{V}} &=-\Omega_\times\Tilde{V} + \frac{1}{m}F+R^T(\mu_w-\alpha_T)
    \end{cases}
    \label{sect5:eq:image_dynamics_system}
\end{equation}
where $\alpha_T$ denotes the linear acceleration of the target plane, i.e. $\alpha_T = \dot{v}_T$.
\begin{theorem}
    Consider the closed-loop system of \eqref{sect5:eq:image_dynamics_system} with the control law
    \begin{equation}
        \frac{1}{m}F = -\pi_\eta k_1(k_2W-q)+k_3\eta(w_d-\eta^TW)+\mu_m
        \label{sect5:eq:landing_controller}
    \end{equation}
where $(k_1, k_2, k_3)$ are positive scalar gains, $w_d>0$, and $\mu_m$ accounts for measurement noise, which is assumed to be bounded, $\eta$ the target plane direction, $q$ the landmark centroid vector in the camera frame and $W$ the translational optical flow. If the condition
\begin{equation}
    k_3w_d >\text{max}\{|\eta^T\mu(t)|\}+\epsilon
    \label{sect5:eq:condition_convergence}
\end{equation}
with $\epsilon>0$ arbitrarily small, and $\mu$ the bounded disturbance vector defined as
\begin{equation}
    \mu:=\mu_m+R^T(\mu_w-\alpha_T),
\end{equation}
is satisfied for all $t\geq0$, then for any initial condition such that $d_0 :=d(0)>0$, the solutions to the system exist for all $t\geq 0$ and 
\begin{enumerate}
    \item[1)] the height $d(t)$ is positive and bounded for all $t\geq 0$ and converges exponentially to zero;
    \item[2)] the state $(\mathcal{C}(t), \Tilde{V}(t)$ is bounded and converges exponentially to zero.
\end{enumerate}
The control law \eqref{sect5:eq:landing_controller} solves problem \ref{sect5:problem:image_dynamics_system}.
\end{theorem}
Refer to \cite{Serra:2016:Primeiroqueli} for a detailed proof of the convergence of this system. 
\section{Simulation Architecture}
\begin{figure}[H]
    \centering
    \includegraphics[scale=0.14]{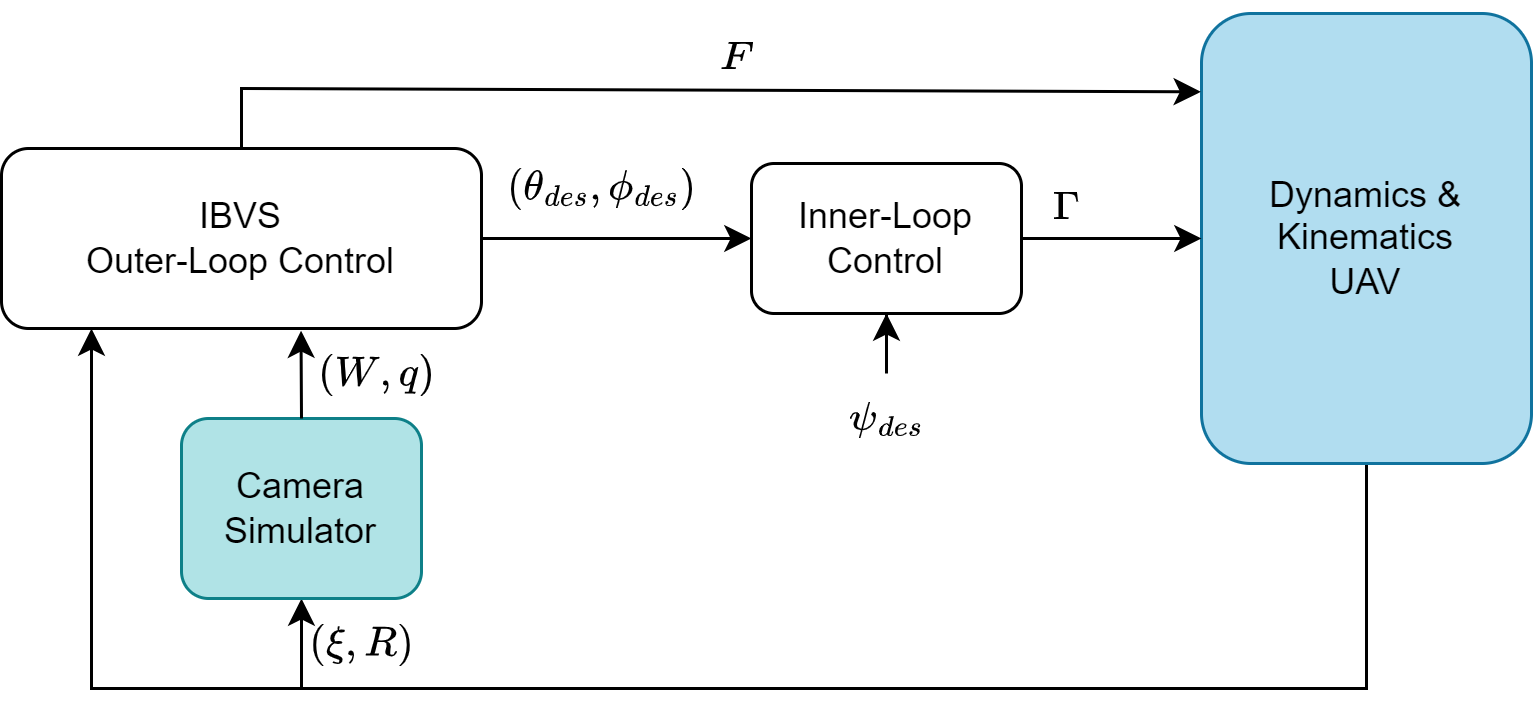}
    \caption{MATLAB simulation scheme}
    \label{ch8:fig_matlab_scheme}
\end{figure}

The image features associated with the centroid vector, defined in \eqref{eq: centroid_vector_q}, represent a set of significant landmarks within the image and the translational optical flow is computed from \eqref{eq: translational_optical_flow}. The initial selection of image points is based on the desired area of integration, and the Lucas-Kanade algorithm is employed to compute the optical flow for these selected image points.
To evaluate the performance of the proposed control law, a series of simulations were carried out using both MATLAB \cite{Matlab:website} and Gazebo 3D software environment \cite{Gazebo}. The proposed control algorithm is represented in Figure \ref{ch8:fig_matlab_scheme}. The camera simulator generates the image acquired by the aerial vehicle's camera at a fixed frequency of 30 frames per second and every time that each new frame is generated, the employed visual features of the IBVS controller, $W$ and $q$, are updated and the control algorithm is applied. 

\begin{figure}[H]
    \centering
    {\includegraphics[trim = 0cm 0cm 0cm 1.35cm, clip, scale=0.5]{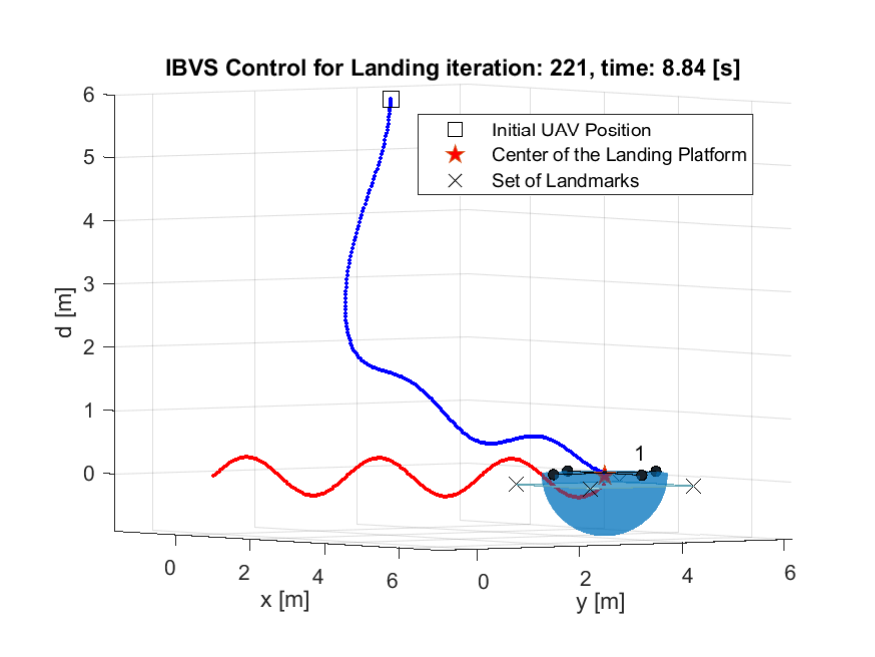}}
    \caption{Simulation of a trajectory in 3-D space $(x, y, d)$ }
    \label{ch7:fig_trajectories-3d_example}
\end{figure}

Figure \ref{ch7:fig_trajectories-3d_example} depicts a trajectory in 3-D space performed by the aerial vehicle and the landing target. The relevant landmarks of the target are also plotted with cross markers, and its center is represented by the red star symbol.

\begin{figure}[h]
    \centering
    \includegraphics[trim = 1.6cm 1.4cm 1.7cm 1.6cm, clip, scale=0.5]{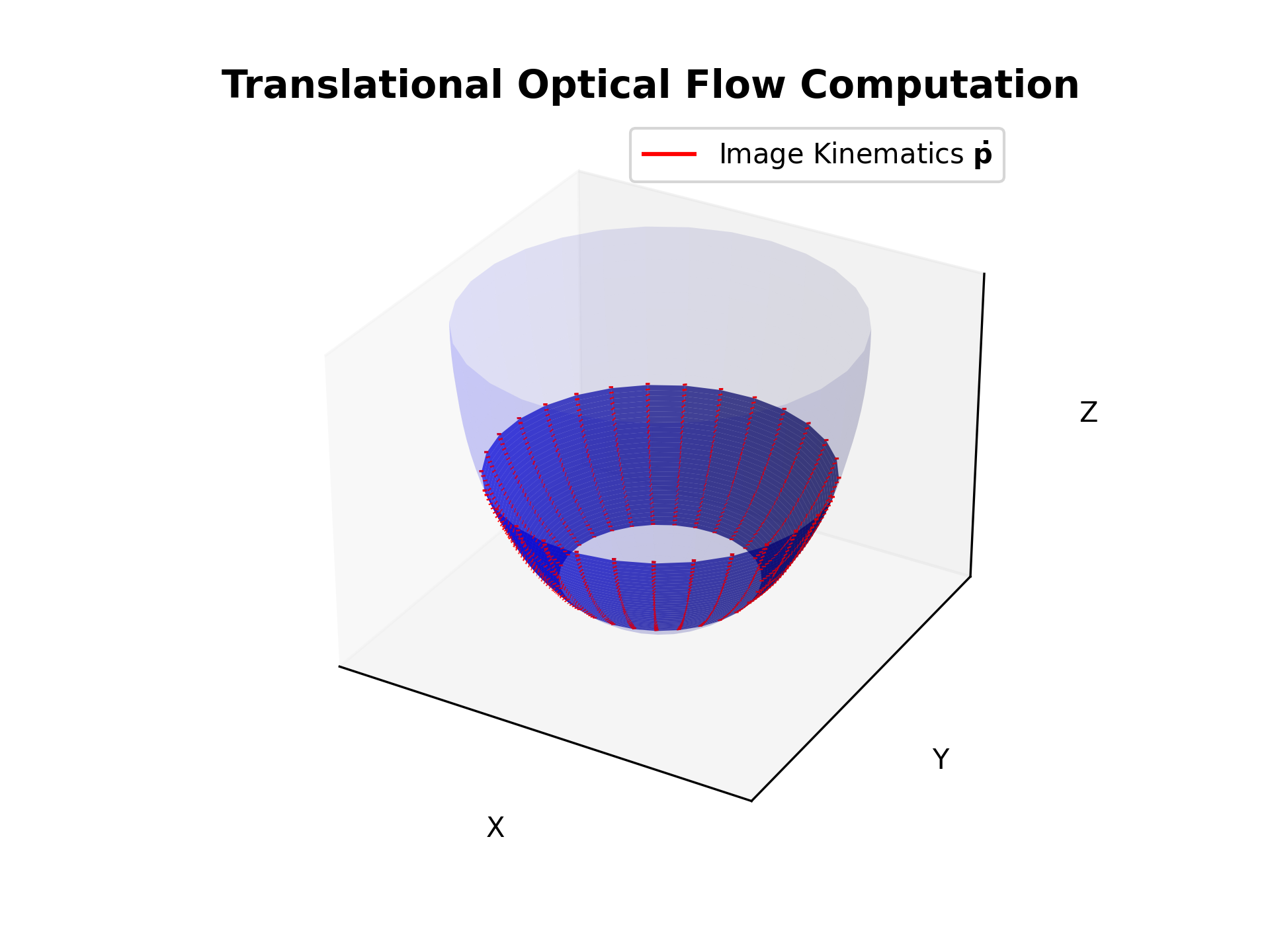}
    \caption[Visual representation of a region of the target plane in spherical coordinates]{Visual representation of a region of the target plane in spherical image geometry with $\theta_0 = \pi/6$ and $\theta_1 = \pi/3$}
    \label{fig:ch7_flow_field_spherical}
\end{figure}

The approach of transitioning to a more realistic simulation environment in Gazebo serves as a necessary intermediate step in the process of implementing the algorithms in an actual quadrotor. More precisely, the simulation process is synchronized with a Software-In-The-Loop (SITL) PX4 setup \cite{Furrer2016}, simplifying the eventual integration with actual PX4 hardware. Given that the main goal is landing an UAV on a moving target, it was necessary to develop an interface to transmit in real-time the captured images from the camera attached to the quadrotor through the Gazebo GStreamer \cite{gstreamer:website} UDP connection to the algorithm in the local computer. The translational optical flow $W$ is computed based on the optical flow field provided by a pyramidal implementation of the Lucas-Kanade algorithm. As described in section \ref{section:w_computation}, the flow field is computed for the 2-D image, and then projected onto a spherical camera, from which the resulting average of the projected vectors is used to compute the translational optical flow. See Figure \ref{fig:ch7_flow_field_spherical} for a visual representation.

\begin{figure}[h]
    \centering
    \includegraphics[scale=0.45]{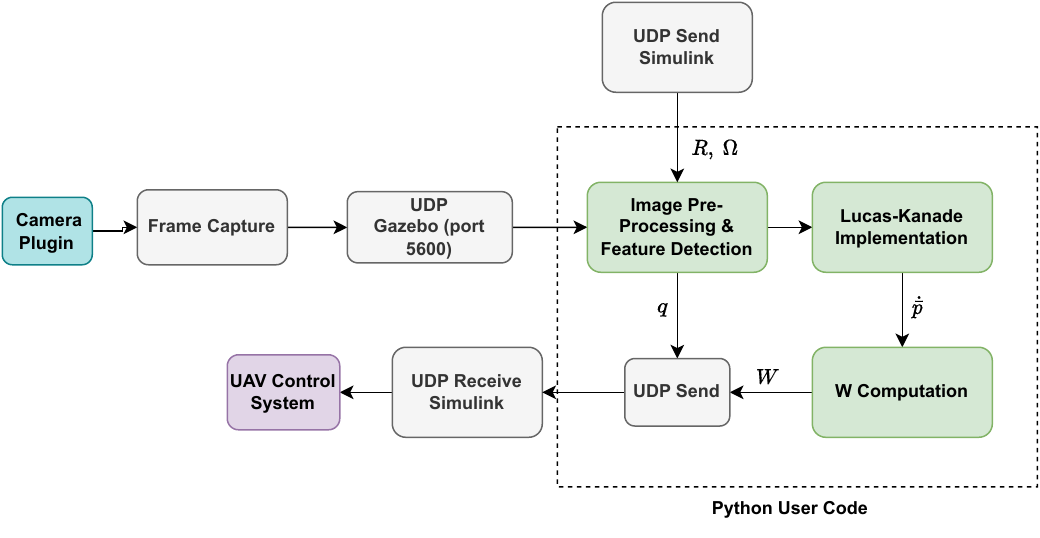}
    \caption{Image processing diagram from Gazebo simulation setup}
    \label{ch7:fig:image_process}
\end{figure}

The local computer establishes two UDP connections with MATLAB: one to provide the visual measurements obtained from the image to the outer-loop controller and another one to receive the inertial data (angular velocity and orientation) of the aerial vehicle (Figure \ref{ch7:fig:image_process}). 

\begin{figure*}[!t]
\centering
\subfloat[]{\includegraphics[width=2.1in]{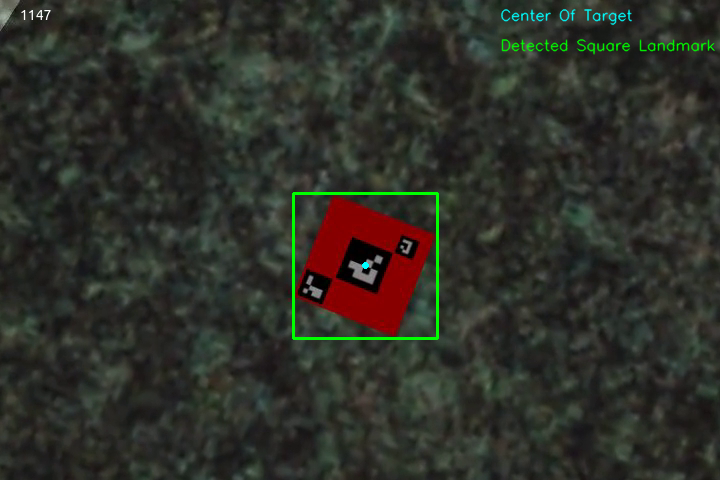}%
\label{fig_first_case}}
\hfil
\subfloat[]{\includegraphics[width=2.1in]{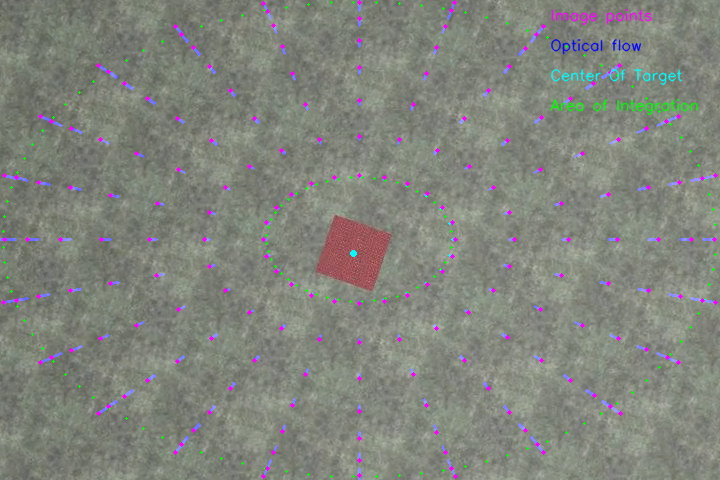}%
\label{fig_second_case}}
\caption{Camera view point of a landing manoeuvre. The cyan dot corresponds to the center of the target; the pink dots correspond to the detected planar image points and the blue arrows denote the estimated flow field in the image plane}
        \label{ch7: example_algorithm}
\end{figure*}

To optimize optical flow detection from the image frames, different textured planes were used and a red-colored textured target with ARuco markers \cite{ARTags} was created. The corners of the red-colored square target are detected on the basis of the HSV (hue, saturation, value) and are considered the $p_i$ landmarks under spherical projection for the proposed work. Note that the ARuco markers present on the target, although not currently used, can be detected to ensure that $q$ is computed in scenarios where there is vision occlusion or errors in landmark detection during the evolution of the system. Figure \ref{ch7: example_algorithm} presents an example of the camera's point of view during the landing operation in the simulation environment.

Synchronizing the inertial data provided by the PX4 SITL and the camera images can prove to be challenging, especially when the quadrotor is under high angular velocities (and the orientation of the aerial vehicle is changing very fast). Moreover, several tasks were limited by the computing capabilities of the local computer/network, such as real-time data transmission and the computation of translational optical flow. In this work, we were able to run the algorithm at a rate of 150 Hz; nevertheless, we encountered significant issues resulting from unpredictable delays within the Gazebo simulation environment and camera streamer. In a real-world setup with a physical quadrotor, many of these challenges are notably less pronounced or even non existent. Data synchronization is more straightforward, as physical sensors provide real-time data without the complications of a simulated environment. In addition, the effort required to process images and implement control algorithms is typically much lower than that required to simulate the entire virtual system and generate images. Real-world setups, while presenting their own challenges, deal with the physical environment directly, reducing the overall computational burden and are less prone to unexpected delays.

\section{Results}
\subsection{Landing on a Moving Target with Oscillatory Motion in the $Z$-axis (MATLAB)}
\begin{figure}[htp]
\captionsetup{farskip=0pt,nearskip=4pt}
\subfloat[Position $\xi(t)$ and $\xi_T(t)$]{\includegraphics[width=0.5\columnwidth]{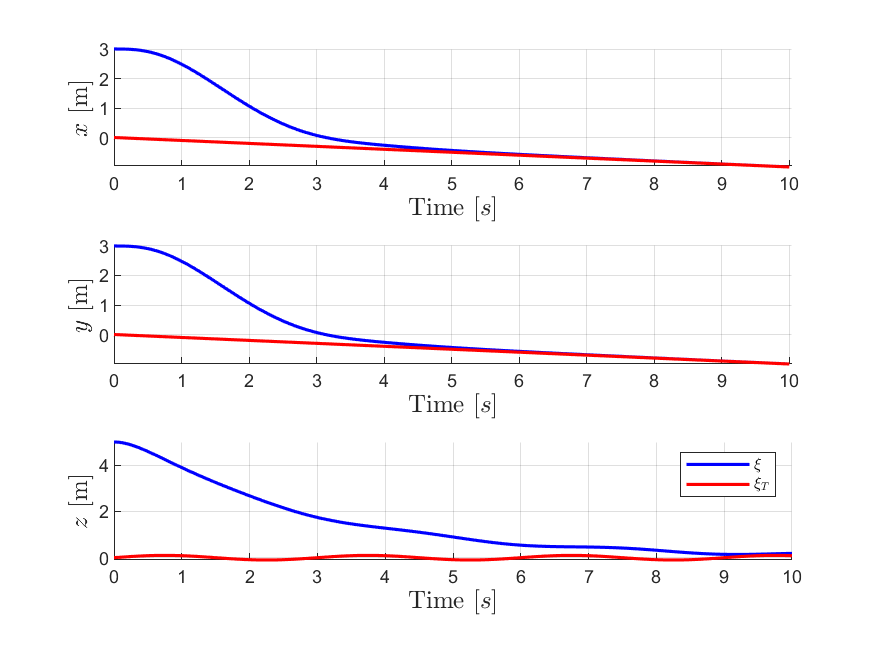}}%
\subfloat[Vector $q(t) = \frac{\mathcal{C}(t)}{d(t)}$ ]{\includegraphics[width=0.5\columnwidth]{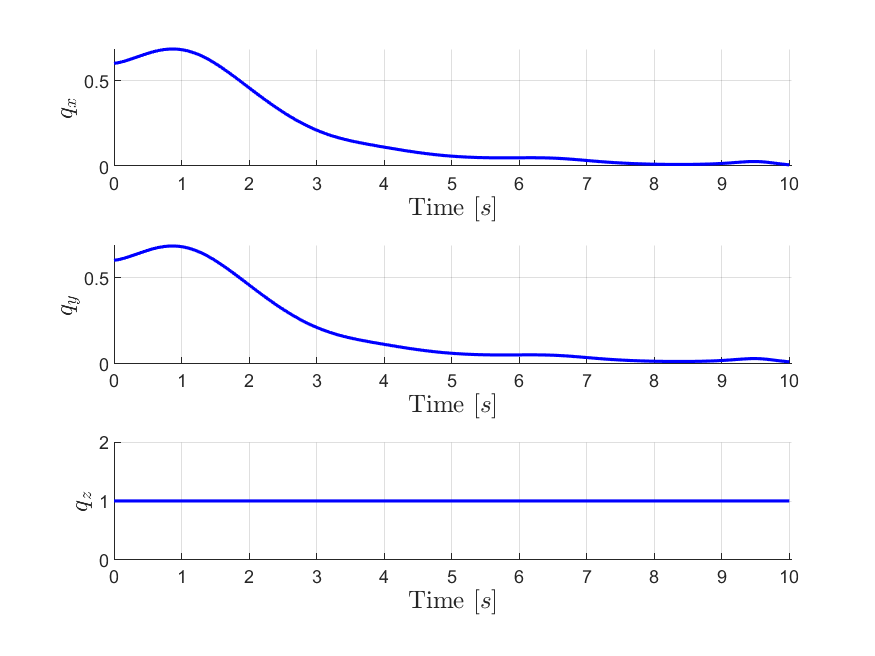}}
\caption{Time evolution of position measurements. The red lines correspond to the target's position}\label{ch8:fig:oscillating_position}
\end{figure}

\begin{figure}[htp]
\captionsetup{farskip=0pt,nearskip=4pt}
\subfloat[Velocity $V(t)$ and $V_T(t)$]{\includegraphics[width=0.5\columnwidth]{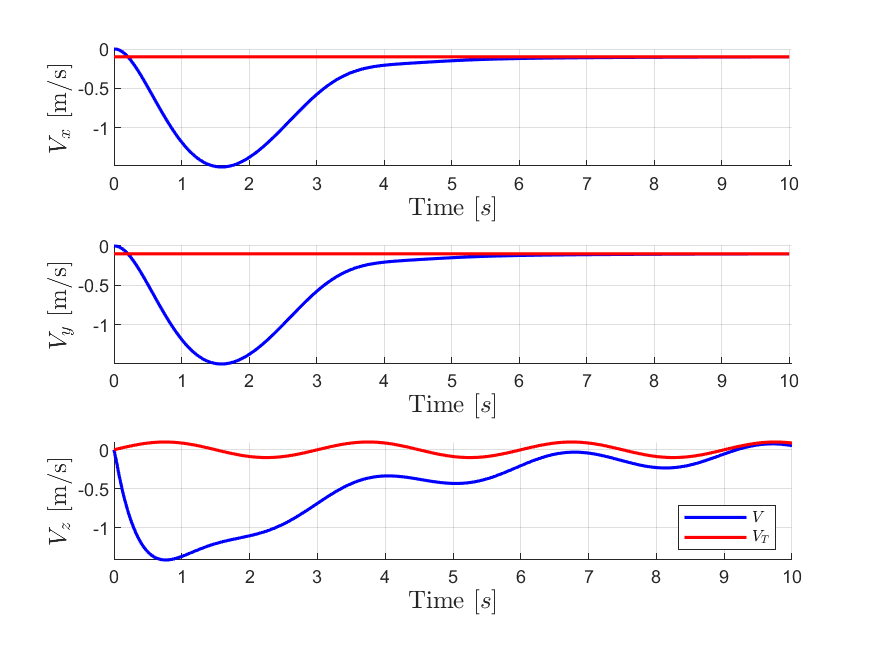}}%
\subfloat[$W(t)=\frac{V(t)-V_T(t)}{d(t)}$]{\includegraphics[width=0.5\columnwidth]{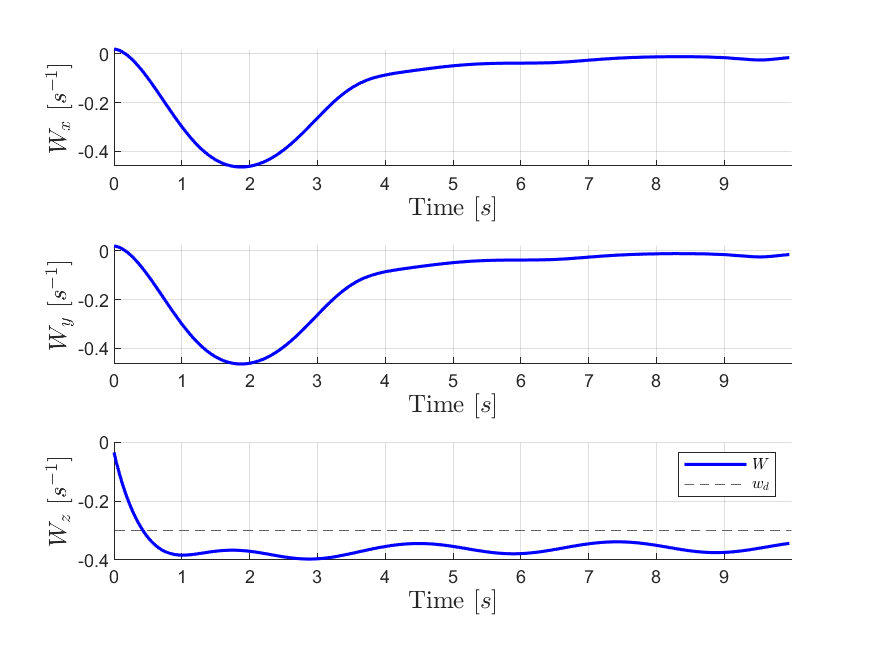}}
\caption{Time evolution of velocity measurements. The red lines correspond to the target's velocity }\label{ch8:fig:oscillating_velocity}
\end{figure}

We start by presenting simulations in MATLAB to validate the controller in perfect conditions. Here, both the quadrotor aerial vehicle and the sensor systems are ideal and serve as a useful starting point for providing a comprehensive understanding of the vision-based controller's performance under error-free measurements. The first application scenario is a moving target with a velocity profile of $V_T(t) = [-0.1\ -0.1\ 0.1sin(\frac{2\pi t}{3})]^T\ ms^{-1}$. The initial position and velocity of the quadrotor is set at $\xi(0) = [3\ 3\ 5]^T\ m$ and $V(0)=[0\ 0\ 0]^T \ ms^{-1}$, respectively. From close examination of Figure \ref{ch8:fig:oscillating_position} and Figure \ref{ch8:fig:oscillating_velocity}, we observe that the position of the aerial vehicle approximates to the center of the target, whereas the relative velocity approaches zero. These results satisfy the convergence theorem (\ref{sect5:eq:condition_convergence}) and demonstrate that the vision-based controller exhibits good performance converging to the desired objective.

\subsection{Stochastically Moving Target (MATLAB)}
\label{ch8:results_stochastically}
\begin{figure}[h]
    \centering
    \includegraphics[scale=0.5]{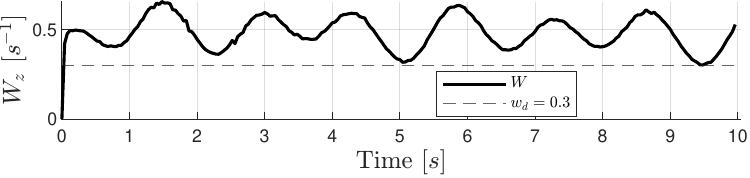}
    \caption[Vertical translational optical flow, simulation of a stochastically moving platform]{Vertical translational optical flow $-W_z(t)$}
    \label{ch8:fig:stoch_wz}
\end{figure}

\begin{figure}[htp]
\centering
\captionsetup{farskip=0pt,nearskip=4pt}
\subfloat[Height $d(t)$]{\includegraphics[trim = 4cm 9cm 4cm 9cm, clip, width=0.45\columnwidth]{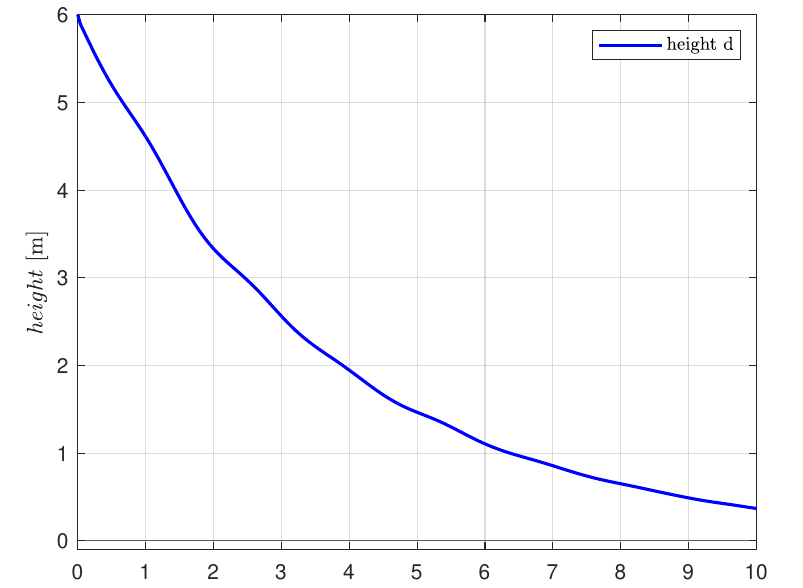}}%
\subfloat[Position $z(t)$]{\includegraphics[trim = 4cm 9cm 4cm 9cm, clip, width=0.45\columnwidth]{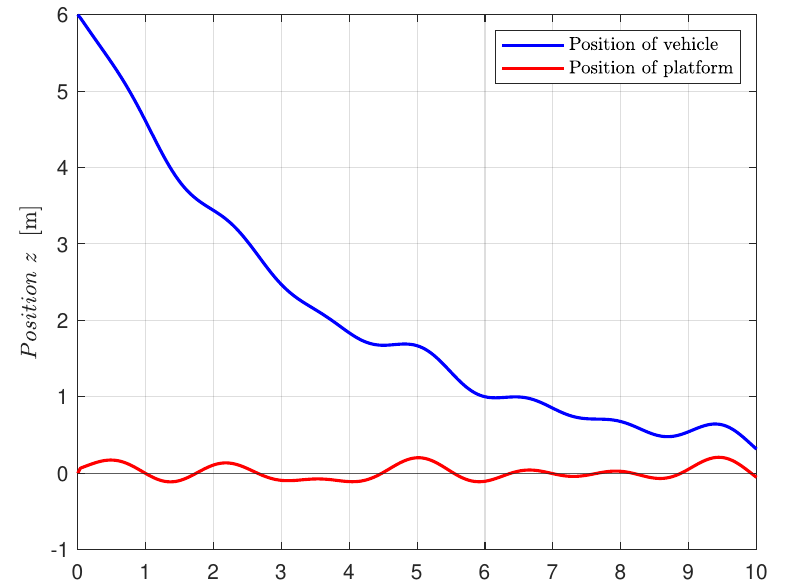}}
\caption{Simulation of vertical landing on a stochastically moving platform}\label{ch8:fig:stoch_simulation_position}
\end{figure}

In the next simulation, we present results regarding a stochastically moving target, and guarantee that the vision-based controller performs well under aperiodic and time-varying vertical motions. The target velocity emulates the landing on a platform that is moving forward with a constant speed and variable up and down oscillatory profile. A practical application that can roughly be described by this velocity profile is the landing on a floating vessel that is moving forward and suffering the influence of sea waves. 
Figure \ref{ch8:fig:stoch_wz} presents the vertical translational optical flow and Figure \ref{ch8:fig:stoch_simulation_position} shows the closed-loop trajectory of the vertical motion of the system. Note that, during the simulation, the relative velocity $\Tilde{V}$ and the height $d$ are assumed to be unknown and are not explicitly applied in the control system. 

\subsection{Landing on a Moving Target Using Sensor Measurements (Gazebo)}
\begin{figure}[htp]
\captionsetup{farskip=0pt,nearskip=4pt}
\subfloat[Position]{\includegraphics[width=0.5\columnwidth]{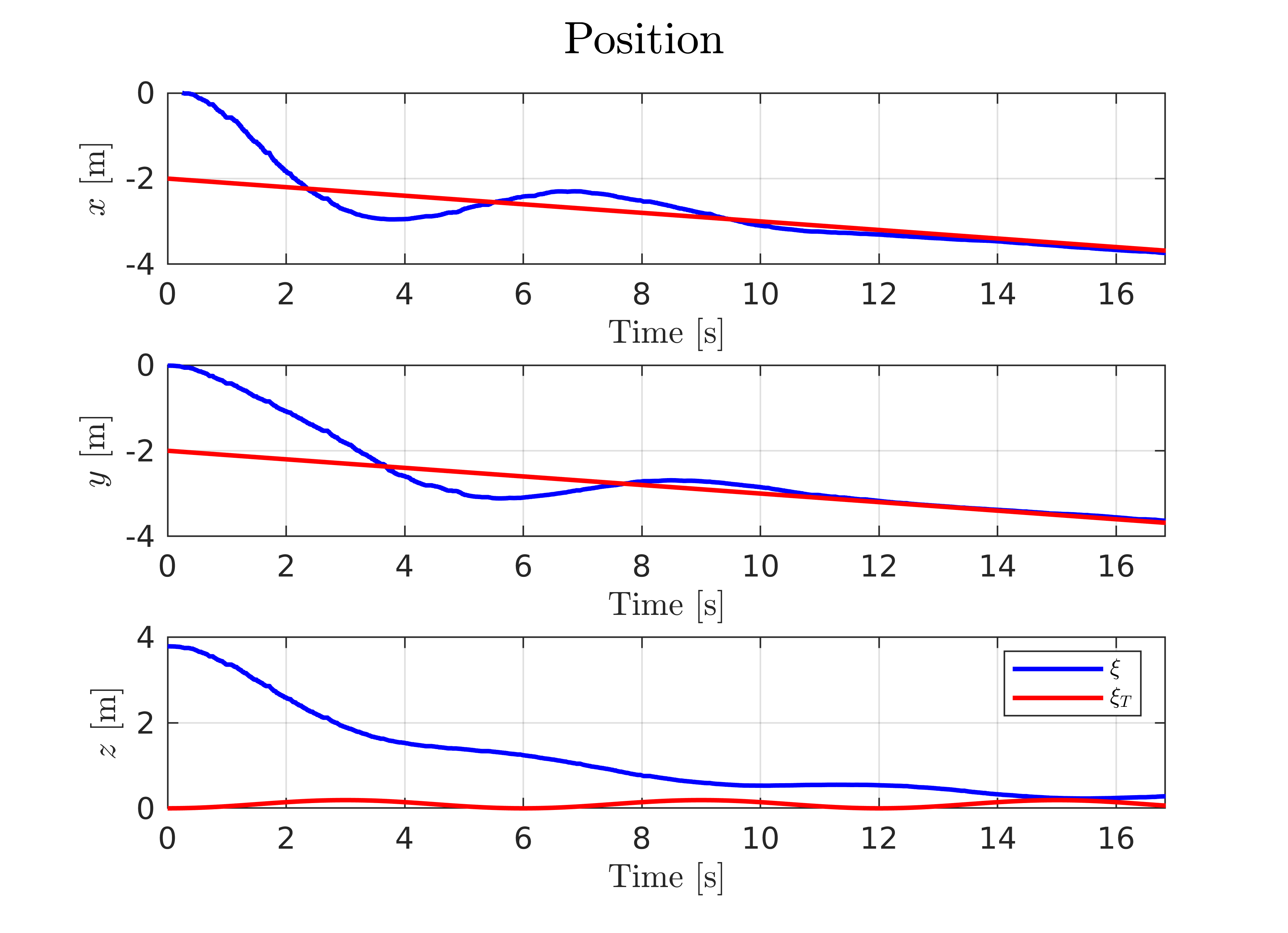}}%
\subfloat[Velocity]{\includegraphics[width=0.5\columnwidth]{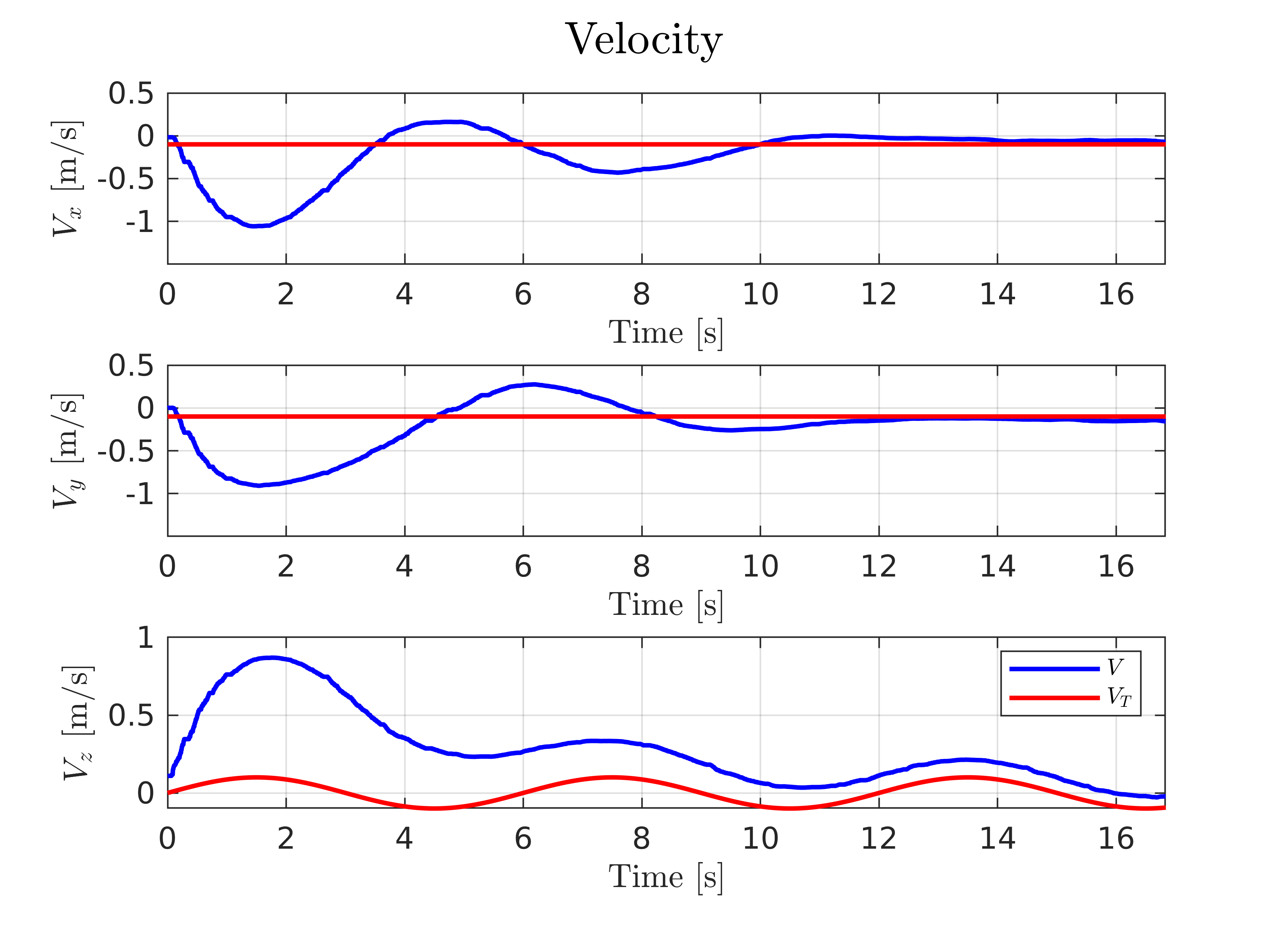}}
\caption{Time evolution of position and velocity. The red lines correspond to the target and the blue lines correspond to the quadrotor. The initial position of the landing platform is $\xi_T(0) = [-2\ -2\ 0]\ m$}\label{ch8:fig:sensor_oscillatory_pos_vel}
\end{figure}

For the next experiment, we present simulations in Gazebo where the outer-loop controller developed for the quadrotor was tested. The target is moving with a velocity of $V_T(t) = [-0.1\ -0.1\ -0.1sin(\frac{2\pi t}{6})]^T\ ms^{-1}$ and the visual position measurement $q = \frac{\mathcal{C}}{d}$ and visual velocity measurement $W=\frac{V-V_T}{d}$ are both provided by the real IMU and GPS system included in the simulation environment. 
Figure \ref{ch8:fig:sensor_oscillatory_pos_vel} shows the time evolution of the position and velocity of the system. 

\begin{figure}[htp]
\captionsetup{farskip=0pt,nearskip=4pt}
\subfloat[$W(t)$]{\includegraphics[width=0.5\columnwidth]{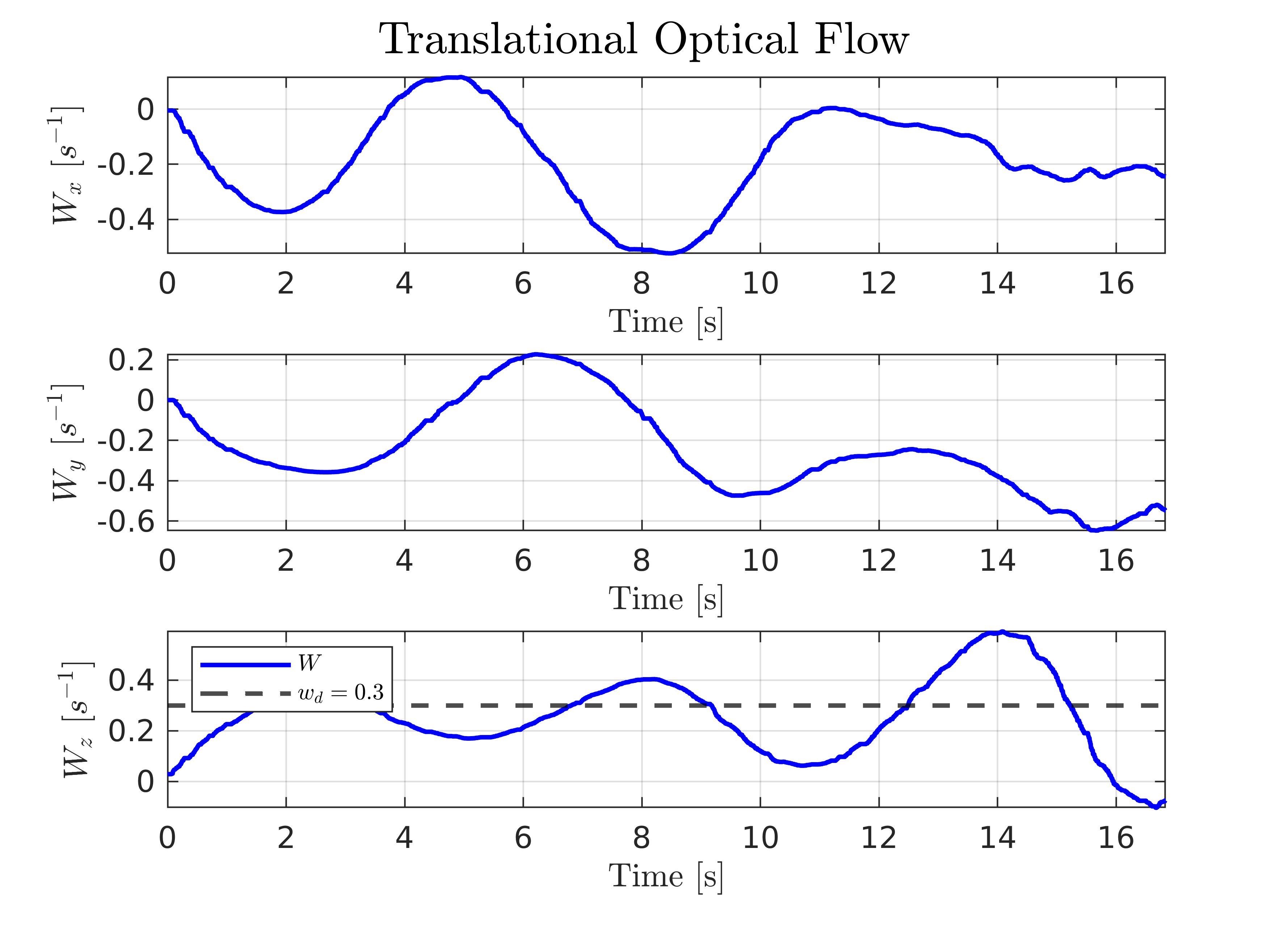}}%
\subfloat[$q(t)$]{\includegraphics[width=0.5\columnwidth]{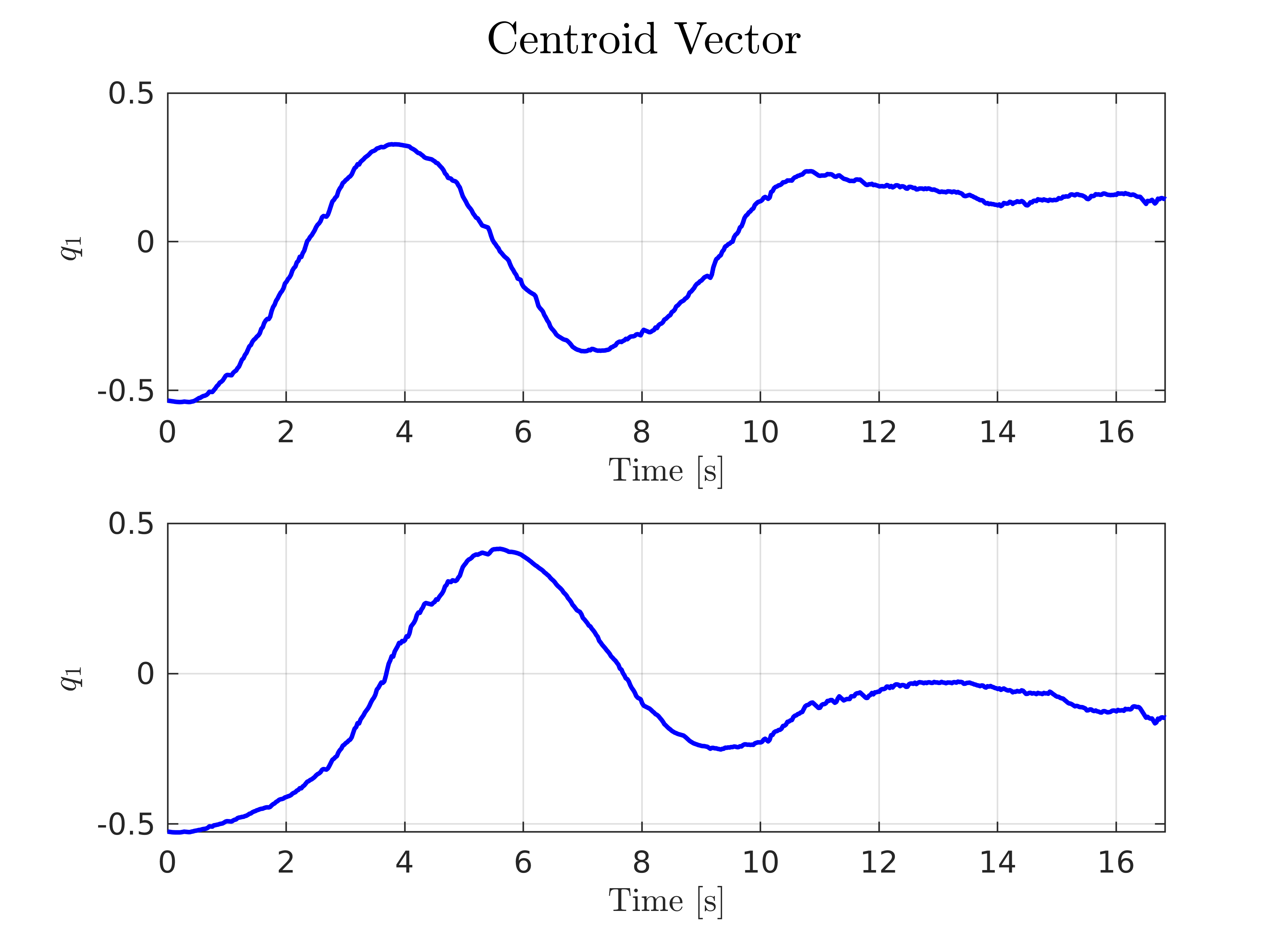}}
\caption{Time evolution of the measured visual information. The dashed lines correspond to the desired translational optical flow $w_d$}\label{ch8:fig:sensor_oscillatory_visual_measurements}
\end{figure}

Figure \ref{ch8:fig:sensor_oscillatory_visual_measurements} presents the time evolution of the visual measurements $W$ and $q$. 
From close examination of Figure \ref{ch8:fig:sensor_oscillatory_pos_vel} and Figure \ref{ch8:fig:sensor_oscillatory_visual_measurements}, the quadrotor converges to the desired moving target and the vision-based controller ensures that the aerial vehicle achieves a smooth touchdown, keeping at all times a positive distance at all times, $d(t)>0$. Notice also that the relative velocity $\Tilde{V}(t) = V(t)-V_T(t)$ converges to zero.

\subsection{Translational Optical Flow and Target Detection (Gazebo)}

\begin{figure}[htp]
\captionsetup{farskip=0pt,nearskip=4pt}
\subfloat[$W(t)$]{\includegraphics[width=0.5\columnwidth]{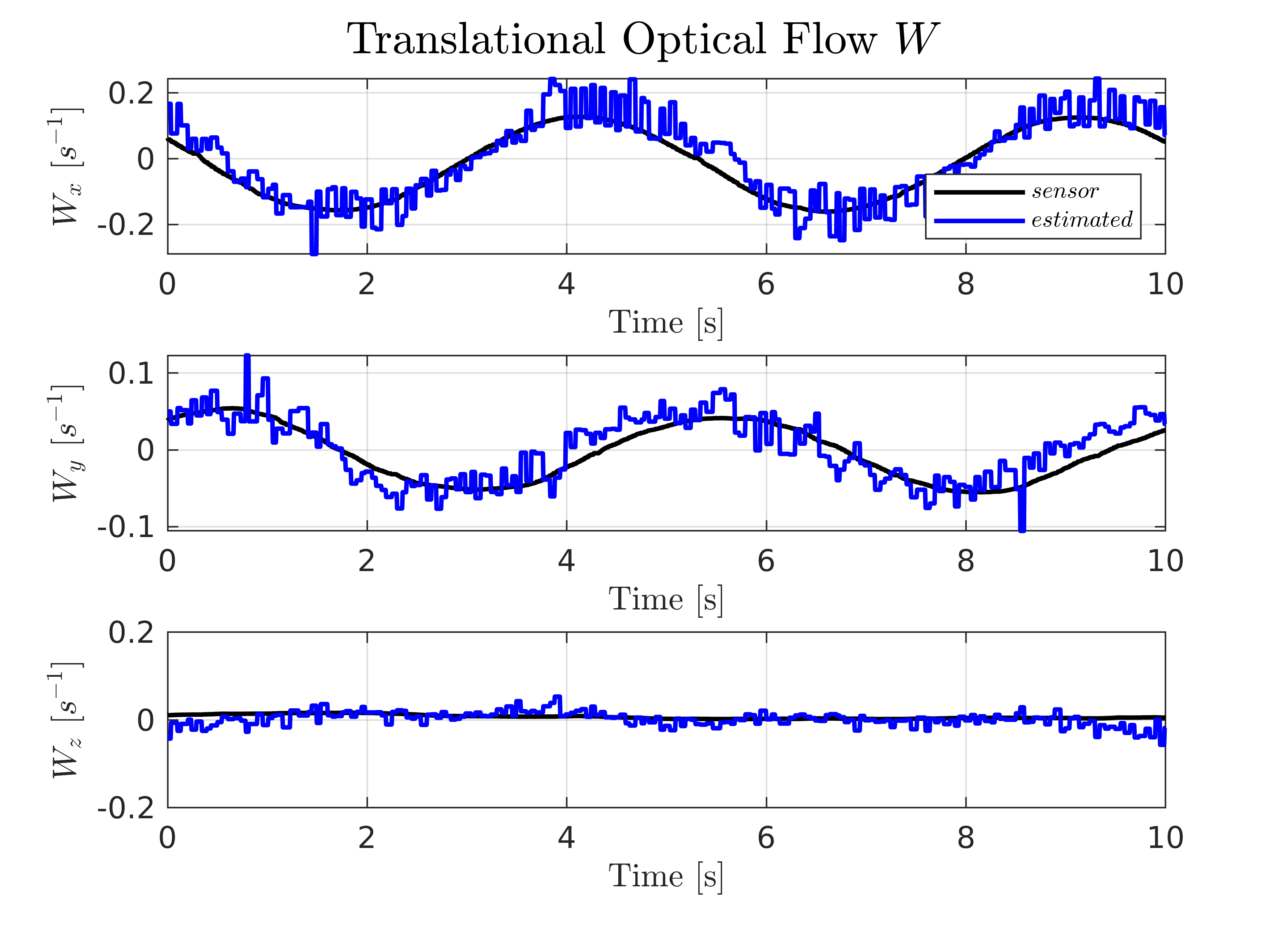}}%
\subfloat[$V(t)$]{\includegraphics[width=0.5\columnwidth]{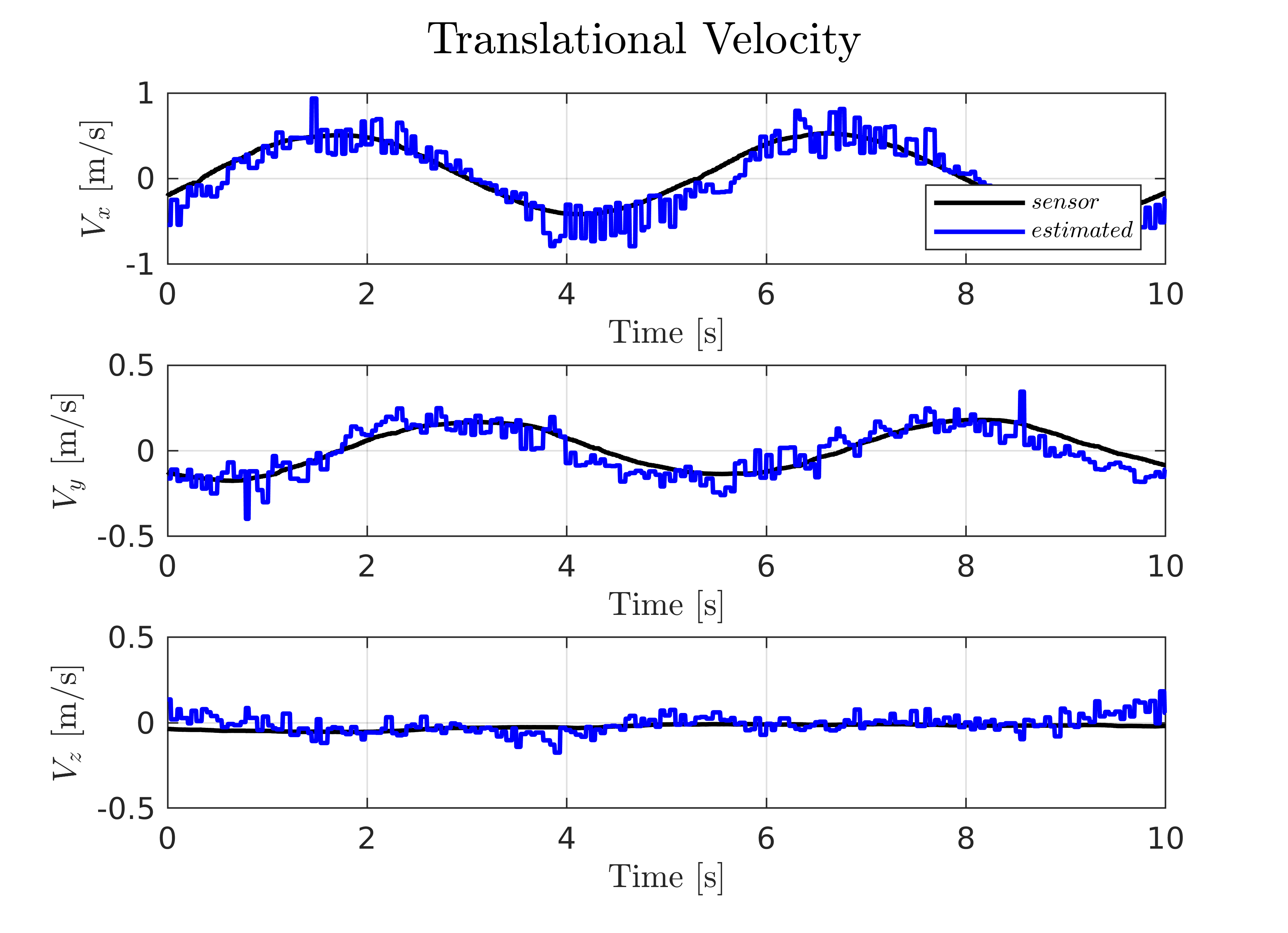}}
\caption{Time evolution of the translational movement of the quadrotor over a horizontal sinusoidal curve at a height of approximately $d =4\ m$}\label{ch8:fig:derotation_w}
\end{figure}

For the next simulation, we control the quadrotor to follow a horizontal sinusoidal curve at a height of $d=4\ m$ in order to analyze the cancellation of the ego-motion of the quadrotor \eqref{eq:derotation} and ensure that only the translational optical flow is calculated from the image frames. The obstacle introduced by rotations is that the optical flow algorithm cannot distinguish translational and rotational movements from images alone. This occurs because in terms of the detected image features by the camera, the rotational movement is the same as the translational one. Thus, it is necessary to obtain in real time both the inertial data and the image frames to obtain the translational component of the average optical flow $\phi_w$. In this simulation, the optical flow is computed with $N=300$ points and the window is $15^\circ$ by $75^\circ$ around $\eta$. As depicted in Figure \ref{ch8:fig:derotation_w}, the computation of the translational optical flow is visibly noisy, but accurately estimates the translational movement between the quadrotor and the environment.

\begin{figure}[h]
    \centering
    \includegraphics[scale=0.35]{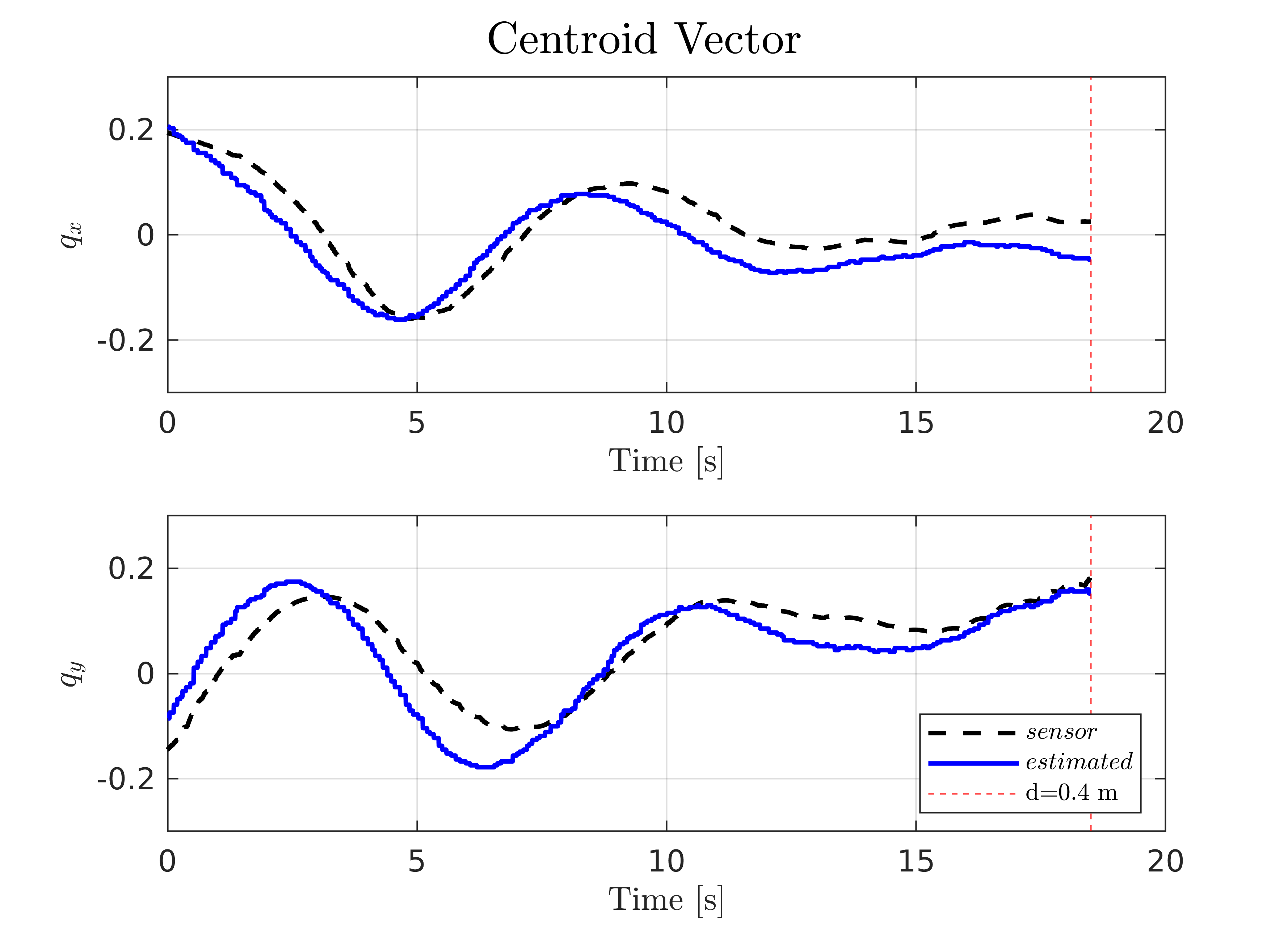}
    \caption{Time evolution of the centroid vector $q$ in the camera frame}
    \label{ch8:fig:gazebo_centroid_vector_estimation}
\end{figure}

\begin{remark}
    In the scenario of Figure \eqref{ch8:fig:derotation_w}, the translational optical flow computed from the images only accounts for the quadrotor's velocity and position, as the target below the camera is static and positioned at the origin. Hence, we compare $V$, and the estimated translational velocity, $V_{estimated} = W_{estimated}d$, where $d$ is the vertical position of the quadrotor. 
\end{remark}
In the next step, we analyze the estimation of the centroid vector in the camera frame during a landing operation. The target is initially placed at $\xi(0) =[0\ 0\ 0]\ m$ and the initial position of the quadrotor is $\xi_T(0) = [1\ -1\ 5]\ m$. Note that the initial state of the system is constrained by the limited field of view of the camera, and although it can be arbitrary, we need to ensure that the visual position measurement $q$ is continuously detected during the evolution of the system. Figure \ref{ch8:fig:gazebo_centroid_vector_estimation} presents the image feature $q$ during the landing operation. We can observe, although with a slight delay (as a result of the image processing algorithm), the estimated $q$ closely resembles $\frac{\mathcal{C}}{d}$ measured from the sensors and converges to zero as the quadrotor approaches to the center of the target. The motors of the quadrotor are shut down automatically when the height $d$ reaches the threshold of $0.4\ m$. This command is considered, given that the quadrotor's landing gear is in close proximity to making contact with the platform or is on the verge of doing so.

\subsection{Landing on a Moving Target with $Z$-axis Oscillatory Motion using Image and IMU data}

\begin{figure}[htp]
\captionsetup{farskip=0pt,nearskip=4pt}
\subfloat[$\xi(t)$]{\includegraphics[width=0.5\columnwidth]{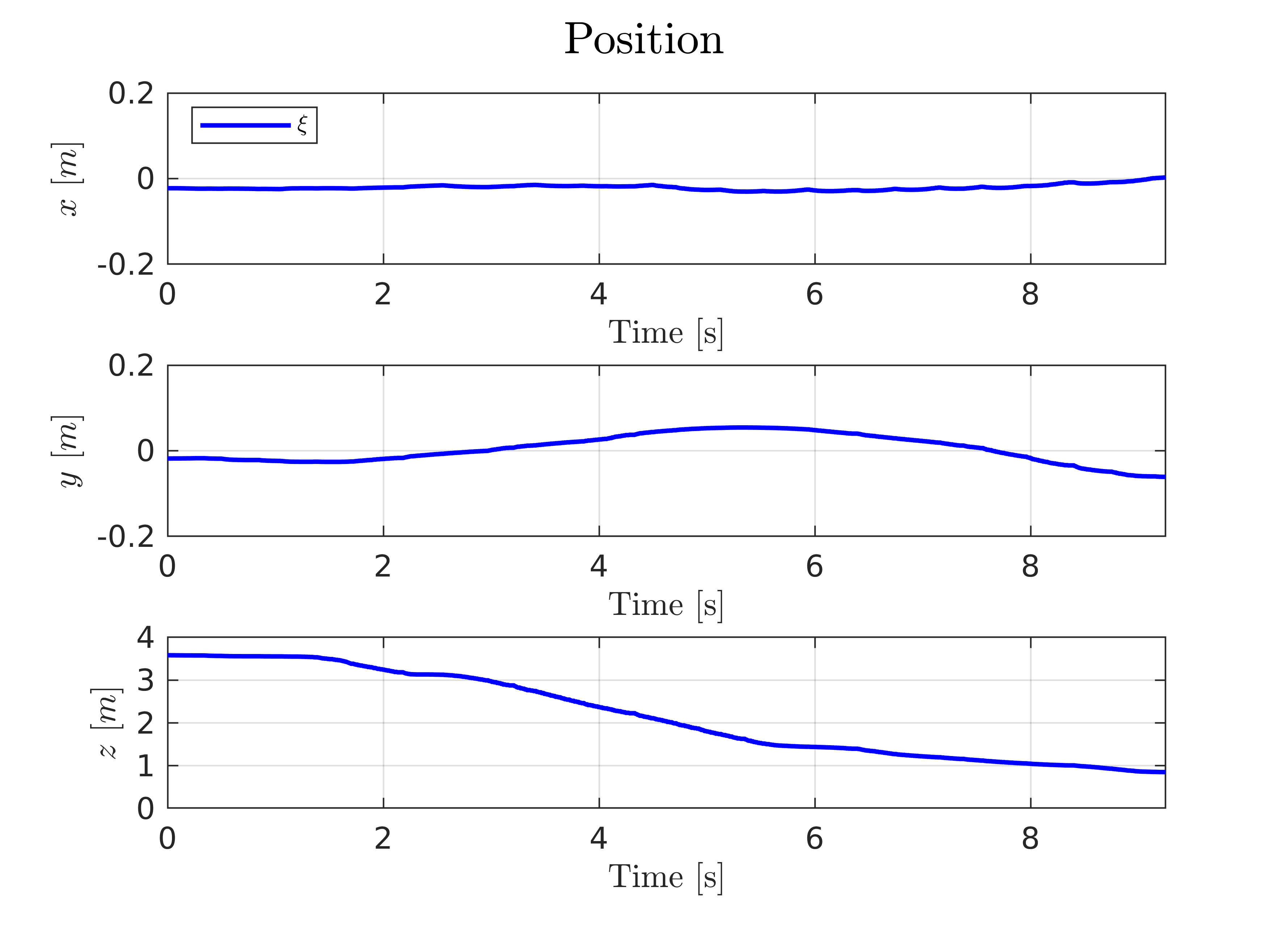}}%
\subfloat[$V(t)$]{\includegraphics[width=0.5\columnwidth]{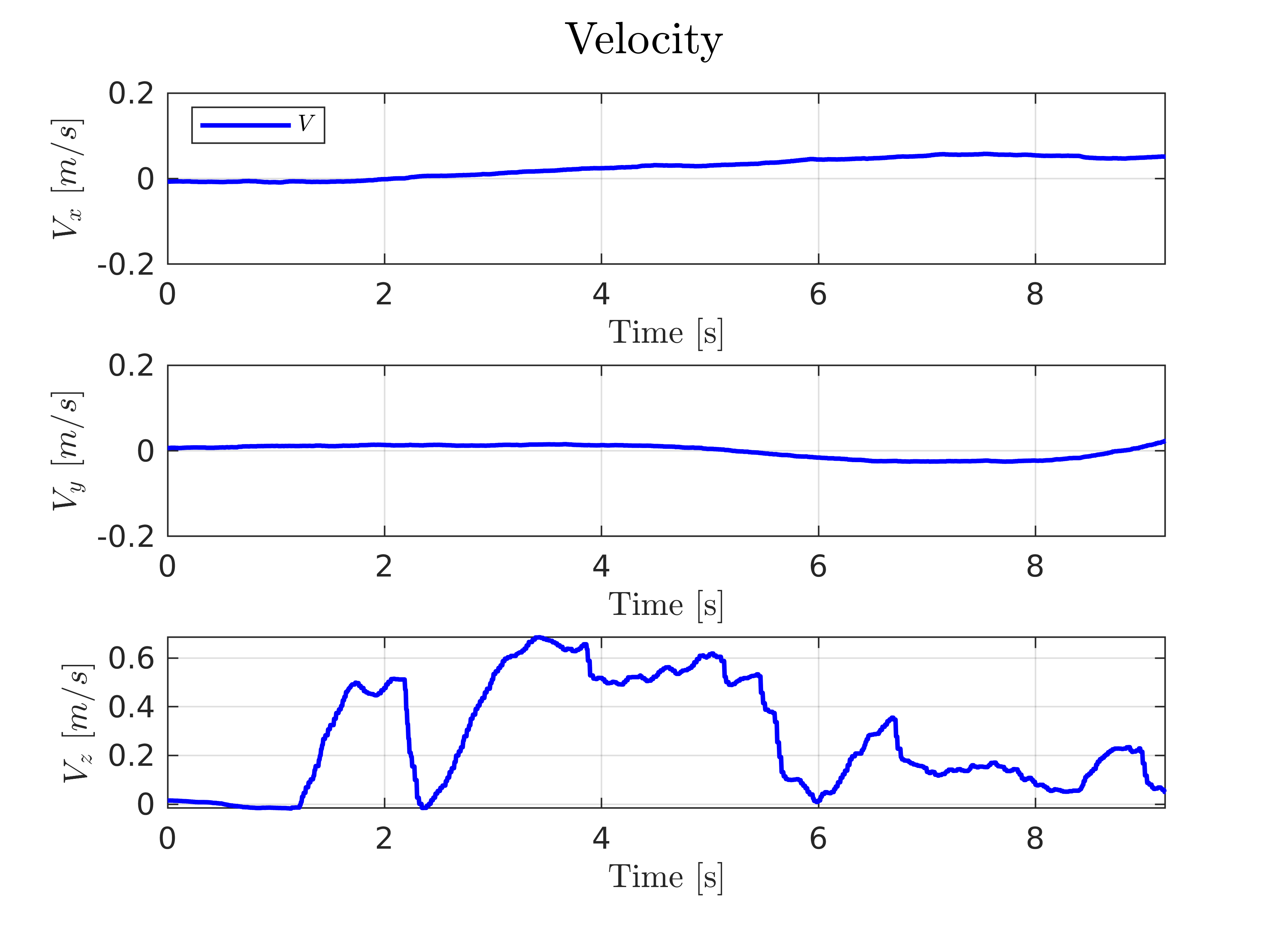}}
\caption{Evolution of position and velocity of the quadrotor in function of time}\label{ch8:fig:gazebo_oscillatory_target}
\end{figure}

\begin{figure}[h]
    \centering
    \includegraphics[scale=0.35]{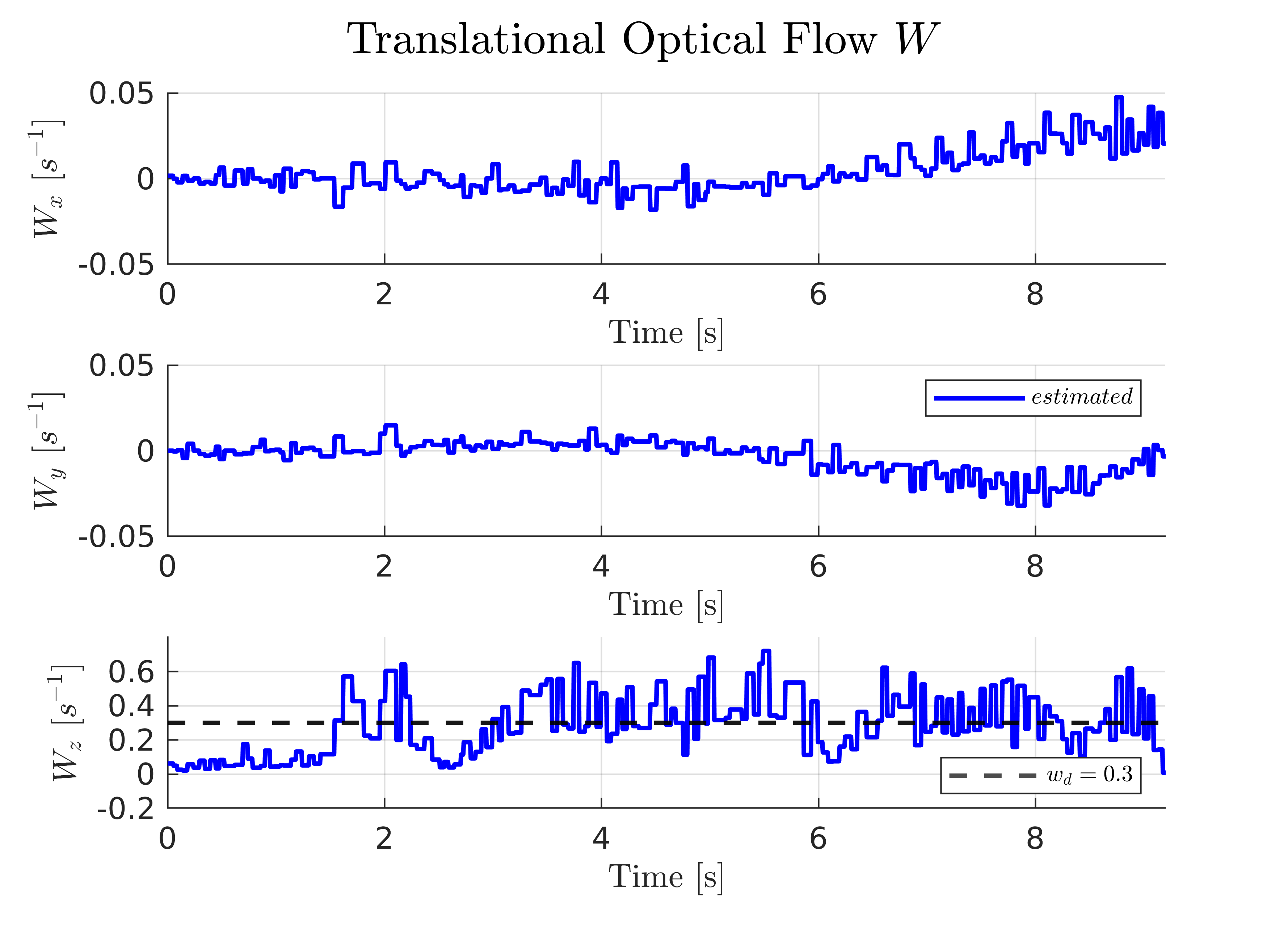}
    \caption{Time evolution of the translational optical flow $W(t)$}
    \label{ch8:fig:gazebo_translational_optical_flow_oscillatory}
\end{figure}

Once it has been shown experimentally that the entire system was able to perform a landing mission using real measurements obtained from IMU and GPS sensors, the next logical step is to test the controller using the visual measurements estimated from the camera image frames. In this experiment, the target is moving with a velocity profile of $V_T(t) = [0\ 0\ -0.1sin(\frac{2\pi t}{6})]^T\ ms^{-1}$ and the initial position of the target is $\xi_T = [0\ 0\ h]^T\ m$, where the height is within the limits of $ h \in \{0,\ 0.1\}\ m$. The quadrotor is set to hover above the target at a height of approximately $d = 4\ m$ and automatically land using the vision-based controller \eqref{sect5:eq:landing_controller} after detecting the observable features of the platform. Note that the position and velocities measured using the sensors embedded in the quadrotor (IMU and GPS) are not used in the control law and the visual feedback are the estimated $q$ and $W$ estimated from image captures. Figure \ref{ch8:fig:gazebo_oscillatory_target} and Figure \ref{ch8:fig:gazebo_translational_optical_flow_oscillatory} presents the simulation results. 

Figure \ref{ch8:fig:gazebo_oscillatory_target} shows the convergence of the position of the quadrotor with the center of the target and it can be observed that the vertical velocity profile of the quadrotor follows similarly to a sinusoidal. Contrary to what was expected, the height $d$ is slowly oscillating during the landing operation as a result of the translational velocity component. This implies that the condition \eqref{sect5:eq:condition_convergence} is not verified in this simulation and therefore, the exponential convergence of the relative velocity and position is not guaranteed. This problem is mainly due to simulation setup constraints (large time latency between inertial measurements and visual features, outer loop command sample rate, non-persistent delays within the simulation environment), which prevents us from choosing a higher magnitude feedback gain $k_3$ which strictly respects the convergence condition. To clarify further, this non ideal behavior of the controller is a result of the aforementioned delays and CPU performance limitations during the simulation. In a real vehicle, these issues would not occur as the computational effort required to process image data and control algorithm implementation is generally much lower than the computational load involved in simulating an entire system and generating images. 

\section{Conclusions}
This dissertation addressed the landing problem of an UAV on a moving target by employing an image-based visual-servo control, using only an onboard camera and an IMU sensor. For control purposes, the centroids of the
images of a collection of landmarks (red corners) for the target are used as position measurement, and the translational optical flow relative to the target is used as velocity measurement, meaning that no direct measurements or explicit estimates of
3-D position or velocity are employed in the control law. The proposed algorithms were implemented in MATLAB and Python, and a simulation interface is developed, replicating the procedures required in a real-life experiment. The proposed vision-based controller is shown to stabilize and guide the UAV to its desired target, avoiding undesired collisions, even in harsh scenarios related to aggressive environmental motions and delays within the simulation environment. 

\subsection{Future Research Directions}
There are some directions in which further work is of interest:
\begin{itemize}
     \item Experimentally test the proposed solution in a real-world system, taking into the consideration the presence of wind disturbances;
     \item Develop filtering methods for the translational optical flow algorithm or apply other methods for estimation, e.g. deep learning models.
\end{itemize}

\vfill

\end{document}